\documentclass{aa}
\usepackage{graphicx}
\usepackage{rotating}
\begin{document}

\newcommand{\as}[2]{$#1''\,\hspace{-1.7mm}.\hspace{.1mm}#2$}
\newcommand{\am}[2]{$#1'\,\hspace{-1.7mm}.\hspace{.0mm}#2$}
\def\approxlt{\lower.2em\hbox{$\buildrel < \over \sim$}}
\def\approxgt{\lower.2em\hbox{$\buildrel > \over \sim$}}
\newcommand{\dgr}{\mbox{$^\circ$}}   
\newcommand{\grd}[2]{\mbox{#1\fdg #2}}
\newcommand{\gsim}{\stackrel{>}{_{\sim}}}
\newcommand{\HI}{\mbox{H\,{\sc i}}}
\newcommand{\HIbf}{\mbox{H\hspace{0.155 em}{\footnotesize \bf I}}}
\newcommand{\HIit}{\mbox{H\hspace{0.155 em}{\footnotesize \it I}}}
\newcommand{\HIsl}{\mbox{H\hspace{0.155 em}{\footnotesize \sl I}}}
\newcommand{\HII}{\mbox{H\,{\sc ii}}}
\newcommand{\IHI}{\mbox{${I}_{HI}$}}
\newcommand{\Jykms}{\mbox{Jy~km~s$^{-1}$}}
\newcommand{\kms}{\mbox{km\,s$^{-1}$}}
\newcommand{\kmsMpc}{\mbox{ km\,s$^{-1}$\,Mpc$^{-1}$}}
\def\lir{{\hbox {$L_{IR}$}}}
\def\lco{{\hbox {$L_{CO}$}}}
\def \ls{\hbox{$L_{\odot}$}}
\newcommand{\LB}{\mbox{$L_{B}$}}
\newcommand{\LBnul}{\mbox{$L_{B}^0$}}
\newcommand{\LBsun}{\mbox{$L_{\odot,B}$}}
\newcommand{\lsim}{\stackrel{<}{_{\sim}}}
\newcommand{\LsunK}{\mbox{$L_{\odot, K}$}}
\newcommand{\LsunB}{\mbox{$L_{\odot, B}$}}
\newcommand{\LsunMsun}{\mbox{$L_{\odot}$/${M}_{\odot}$}}
\newcommand{\LK}{\mbox{$L_K$}}
\newcommand{\LKLB}{\mbox{$L_K$/$L_B$}}
\newcommand{\LKLBnul}{\mbox{$L_K$/$L_{B}^0$}}
\newcommand{\LKLsun}{\mbox{$L_{K}$/$L_{\odot,Bol}$}}
\newcommand{\masq}{\mbox{mag~arcsec$^{-2}$}}
\newcommand{\MHI}{\mbox{${M}_{HI}$}}
\newcommand{\MHILB}{\mbox{$M_{HI}/L_B$}}
\newcommand{\MHILBfr}{\mbox{$\frac{{M}_{HI}}{L_{B}}$}}
\newcommand{\MHILK}{\mbox{$M_{HI}/L_K$}}
\newcommand{\KMS}{\mbox{$\frac{km}{s}$}}
\newcommand{\JYKMS}{\mbox{$\frac{Jy km}{s}$}}
\newcommand{\MHILKfr}{\mbox{$\frac{{M}_{HI}}{L_{K}}$}}
\def \ms{\hbox{$M_{\odot}$}}
\newcommand{\Msun}{\mbox{${M}_\odot$}}
\newcommand{\MsunLsun}{\mbox{${M}_{\odot}$/$L_{\odot,Bol}$}}
\newcommand{\MsunLBsun}{\mbox{${M}_{\odot}$/$L_{\odot,B}$}}
\newcommand{\MsunLKsun}{\mbox{${M}_{\odot}$/$L_{\odot,K}$}}
\newcommand{\MT}{\mbox{${M}_{ T}$}}
\newcommand{\MTLBnul}{\mbox{${M}_{T}$/$L_{B}^0$}}
\newcommand{\MTLBsun}{\mbox{${M}_{T}$/$L_{\odot,B}$}}
\newcommand{\nan}{Nan\c{c}ay}
\newcommand{\tis}[2]{$#1^{s}\,\hspace{-1.7mm}.\hspace{.1mm}#2$}
\newcommand{\Vcor}{\mbox{${V}_{0}$}}
\newcommand{\vhel}{\mbox{$V_{hel}$}}
\newcommand{\VHI}{\mbox{${V}_{HI}$}}
\newcommand{\vrot}{\mbox{$v_{rot}$}}
\def\la{\mathrel{\hbox{\rlap{\hbox{\lower4pt\hbox{$\sim$}}}\hbox{$<$}}}}
\def\ga{\mathrel{\hbox{\rlap{\hbox{\lower4pt\hbox{$\sim$}}}\hbox{$>$}}}}

 \title{A search for Low Surface Brightness galaxies in the near-infrared}

  \subtitle{II. Arecibo H{\large \bf I} line observations}

  \author{D. Monnier Ragaigne\inst{1}, 
          W. van Driel\inst{1},
          K. O'Neil\inst{2},
          S.E. Schneider\inst{3},
          C. Balkowski\inst{1}, 
      \and
          T.H. Jarrett\inst{4}
          } 

  \offprints{W. van Driel}

  \institute{Observatoire de Paris, GEPI, CNRS UMR 8111 and Universit\'e Paris 7, 
              5 place Jules Janssen, F-92195 Meudon Cedex, France \\
             \email{delphine.ragaigne@obspm.fr; wim.vandriel@obspm.fr; 
                    chantal.balkowski@obspm.fr}
       \and
               NRAO, P.O. Box 2, Green Bank, WV 24944, U.S.A. \\
              \email{koneil@nrao.edu}
        \and
             University of Massachusetts, Astronomy Program, 536 LGRC, Amherst, 
             MA 01003, U.S.A. \\
             \email{schneide@messier.astro.umass.edu}
        \and
             IPAC, Caltech, MS 100-22, 770 South Wilson Ave., Pasadena, 
             CA 91125, U.S.A. \\
             \email{jarrett@ipac.caltech.edu}
             }

\date{\it Received 20/2/2002 ; accepted 30/4/2002}

  \abstract{ {\rm
A total of 367 Low Surface Brightness galaxies detected in the 2MASS 
all-sky near-infrared survey have been observed in the 21 cm \HI\ line using 
the Arecibo telescope. All have a $K_s$-band mean central 
surface brightness, measured within a 5$''$ radius, fainter than 18 \masq. 
We present global \HI\ line parameters for the 107 clearly detected objects and 
the 21 marginal detections, as well as upper limits for the undetected objects.
The 107 clear detections comprise 15 previously uncatalogued objects and 36
with a PGC entry only. 
}
   }

 \authorrunning{D. Monnier Ragaigne et al.}
 \titlerunning{A search for LSB galaxies in the near-infrared II.}
 
 \maketitle

\section{Introduction}  
The present paper is part of a series presenting the results of a multi-wavelength
(near-infrared, 21-cm \HI\ line and optical) search for Low Surface Brightness 
(LSB) galaxies using the 2MASS all-sky near-infrared survey. 
For further details on the sample and other
publications in this series, we refer to Monnier Ragaigne et al. (2003a, Paper I). 

\subsection{Low Surface Brightness galaxies}  
Observational bias in the selection of galaxies dates back to Messier and Herschel. 
Galaxies are diffuse objects selected in the presence of a contaminating signal: the
brightness of the night sky. The night sky acts as a filter, which, when 
convolved with the true population of galaxies gives the population of galaxies we observe.
In the past few decades, observations of the local universe have shown the
existence of galaxies well below the surface brightness of the average catalogued galaxy.

At present, the LSBs constitute the least well
known fraction of galaxies: their number density and physical properties 
(luminosity, colours, dynamics) are still quite uncertain. This is, per 
definition, mainly due to the fundamental difficulty in identifying them 
in imaging surveys and in measuring their properties.
In order to further investigate the often baffling properties of the LSB 
class of galaxies we selected a sample of them from the 2MASS database, 
accessing a wavelength domain (the near-infrared) hitherto only scarcely explored 
in the study of LSBs.

There is no unambiguous definition of LSB galaxies, although ones in common use
are based on (1) the mean blue surface brightness within the 25 \masq\ 
isophote, (2) the mean surface brightness within the half-light radius,
or (3) the extrapolated central surface brightness of the disc component
alone after carrying out a disc-bulge decomposition.
For 2MASS galaxies, we used the mean $K_s$-band
magnitude within a fixed aperture to identify a sample of galaxies with
relatively low infrared surface brightness.
Because galaxies typically have $(B-K)\sim 3.5$--4 (see below) we selected
galaxies in which the central surface brightness within a 5$''$ radius circular aperture 
was fainter than 18 \masq in $K_s$. This criterion corresponds
roughly to the disc-component definition of LSB galaxies which have a blue
central surface brightness $\mu_{B_0}>22.0$ \masq. Our sample galaxies can have
even lower disc surface brightness levels if the bulge component is significant, but since 
we average over a fixed angular aperture, we may also include some higher central surface 
brightness sources that are more distant so that the aperture includes more of the disc. 
In this paper, we examine the properties of our sample based on 2MASS and HyperLeda data 
and compare it to optically-selected galaxy samples.

LSBs have remarkable properties which distinguish them from high surface 
brightness spirals, notably:
\begin{itemize}
\item LSBs seem to constitute at least 50\% of the total galaxy population in 
number in the local Universe, which may have strong implications for the 
slope of the galaxy luminosity function, on the baryonic matter density and 
especially on galaxy formation scenarios (O'Neil \& Bothun 2000);
\item LSBs discs are among the less evolved objects in the local universe since 
they have a very low star formation rate (van der Hulst et al. 1993; van Zee et
al. 1997; van den Hoek et al. 2000);
\item LSBs are embedded in dark matter halos which are of lower density and 
more extended than the haloes around High Surface Brightness (HSB) galaxies, 
and they are strongly dominated by Dark Matter at all radii (de Blok et al. 1996; 
McGaugh et al. 2001).
\end{itemize}
The star formation history of LSBs has been the subject of recent debate.
The LSBs best studied in the optical and in the near-infrared are blue
(e.g., Bergvall et al. 1999), indicating a young mean stellar age and/or 
metallicity. 
Morphologically, most studied LSBs have discs, but little spiral structure.
The current massive star formation rates in LSBs are an order of magnitude 
lower than those of HSBs (van Zee et al. 1997); \HI\ observations show that LSBs 
have high gas mass fractions, sometimes exceeding unity 
(Spitzak \& Schneider 1998; McGaugh \& de Blok 1997). All these observations are 
consistent with a 
scenario in which LSBs are relatively unevolved, low mass surface density, 
low metallicity systems with roughly constant star formation rate. However, 
this scenario has difficulty accomodating giant LSBs like Malin 1 
(Bothun et al. 1987).

This study of infrared LSBs was also intended to investigate the possibility
of there being a substantial population of red LSBs like those reported by 
O'Neil et al.~(1997). Follow-up studies
initially indicated that these red LSBs had rapid rotational speeds
exceeding 200 \kms\ and they did not seem to follow the `standard' 
Tully-Fisher relation, in the sense that they appeared to be severely 
underluminous for their total mass (O'Neil et al.~2000). More recent observations
by Chung et al.~(2002) indicate that the rotational properties of these 
red LSBs were mismeasured due to confusion with neighbouring galaxies.
An infrared-selected sample should allow us to identify whether there
is a significant population of very red LSBs.

In the present paper a result of 21-cm \HI\ line observations made at Arecibo is 
given, while in paper I in the series a description of the 2MASS LSB galaxy 
sample selection is given,  21-cm \HI\ line observations from \nan\ will be
presented in Paper III (Monnier Ragaigne et al. 2003b), 
optical $BV\!RI$ CCD surface photometry of a sub-sample of 
35 2MASS LSB galaxies will be presented in paper IV (Monnier Ragaigne et al. 2003c),
models of the star formation history of these, and an analysis of the
full data set will be presented in paper V.
Models of the evolution of the 2MASS galaxies presented in paper IV and of 
other samples of LSB galaxies are presented in Boissier et al. (2003).

The 2MASS LSB galaxies sample selection is described in 
Section 2. Observations and data reduction  are presented in Section 3,  and 
the results in section 4 and the conclusions in section 5.

\subsection{The 2MASS all-sky near-infrared survey}  
The Two Micron All Sky Survey, 2MASS, has imaged the entire celestial sphere in the 
near-infrared J (1.25$\mu$m), H (1.65$\mu$m) and $K_s$ (2.16$\mu$m) bands from 
two dedicated, identical 1.3-meter telescopes. 
The 2MASS data have a 
95\% completeness level in $J$, H and $K_s$ of 15.1, 14.3 and 13.5 mag, 
respectively, for `normal' galaxies (Jarrett et al. 2000a); 
for LSB and blue objects the completeness limits are not yet known). 
The Extended Source Catalog (XSC) will consist of more than 1.4 million galaxies 
brighter than 14th mag at $K_s$ with angular diameters greater than $\sim$6 arcsec.
The photometry includes accurate PSF-derived measures and a variety of circular and 
elliptical aperture measures, fully characterizing both point-like and
extended objects. The position centroids have an astrometric accuracy better than 
$\sim$0.5 arcsec. In addition to tabular information, 2MASS archives full-resolution 
images for each extended object, enabling detailed comparison with other imaging surveys. 
Initial results for galaxies detected by 2MASS are described in several publications 
(Schneider et al. 1997; Jarrett et al. 2000a, 2000b; Hurt et al. 2000). 

Although relatively less deep than some of the dedicated optical imaging surveys
made of LSB galaxies over limited areas of the sky, the 2MASS survey allows 
the detection of LSB galaxies with a central surface brightness in the $K_s$ band 
of $\sim$18-20 \masq, corresponding to about $\sim$22-24 \masq\ in $B$, 
extending over the entire sky. The near-infrared data will be less susceptible 
than optical surveys to the effects of extinction due to dust, both Galactic
and internal to the galaxies.

In the present paper the \HI\ line observations made at Arecibo 
of the sample of 2MASS near-infrared selected LSB galaxies is
presented. The sample selection in described in Section 2. 
The observations and the data reduction are presented in Section 3,
and the results in Section 4. A brief discussion of the results is
given in Section 5. An analysis of these observations, and of others 
made for the study, will be given in a future paper (paper V).

\section{Sample selection}  
We have selected 2MASS LSB galaxies using the following two galaxy search routines.
For a more complete description of the selection procedures we refer to Paper I.
The selected LSB objects lie outside the ZoA 
($|$b$|$$>$10$^{\circ}$), but we have also observed 116 objects
within this zone (see Section 4.6).

$\bullet$\, The first is aimed at selecting relatively high signal-to-noise
low central surface brightness (LCSB) galaxies, 
with a mean central $K_s$ surface brightness in the inner 5$''$ 
radius of $K_5$$\geq$18 mag arcsec$^{-2}$,
among the extended sources picked out from the
survey data by the standard 2MASS algorithms (Jarrett et al. 2000).

$\bullet$\, The second is aimed at finding lower signal-to-noise LSB galaxies among 
those sources which were not recognized as such during the standard
extended source selection described above. 
This requires masking all sources detected by the former method, spatially
smoothing the remaining data and running the extended source recognition 
scheme again on the masked images. 

In order to decide which of these faint, fuzzy 2MASS sources really are 
LSB galaxies, additional data were required like those listed in online 
databases such as NED (NASA Extragalactic Database) [http://nedwww.ipac.caltech.edu], 
LEDA (Lyon-Meudon Extragalactic Database) -- recently incorporated in
HyperLeda [http://www-obs.univ-lyon1.fr/hypercat/] --
and Aladin of the Centre de Donn\'es astronomiques de Strasbourg (CDS) 
[http://aladin.u-strasbg.fr] and by inspecting Digital Sky Survey (DSS)
images. Using this selection procedure, a total of 3,800 candidate 2MASS LSB 
galaxies were found. 

The source selection for our survey was made with the 2MASS working database 
available in late 1999, when work on it was still in full progress.  The sources we observed
at Arecibo contain only 2 ``Small'' sources (with isophotal radii 10$''$$<$$r_{K_{20}}$$<$20$''$, 
see Section 4.1), S721N and S1090O, that are not found in the 
subsequent, more reliable, 2MASS public data releases (see paper I for details).
Their names have been put in parentheses in the Tables.
The sample also contains 2 Faint sources (F1N and F10N)  that are not found in 
subsequent public data releases, which not surprising considering they were detected by 
a dedicated LSB galaxy search method and that they were not included in the working 
database.

Due to constraints in the telescope time scheduling and in the 
declination range accessible with the instrument, the area of the sky
observed at Arecibo for our survey ranges from 20$^h$00$^m$ to
06$^h$30$^m$ in right ascension and between 12$^{\circ}$ and 39$^{\circ}$
in declination. Within this area, we selected for observation,
in order of decreasing priority: 
(1) all objects with known velocity,
(2) objects without known velocity, whether previously catalogued 
or not, in order of decreasing diameter.

\section{Observations and data reduction }  
\subsection{Observations} 
We observed a sample (see Section 2) of 367 2MASS LSB galaxies using the
refurbished 305-m Arecibo Gregorian radio telescope in November 2000 and
January 2001, for a total of 125 hours observing time.
Data were taken with the L-Band Narrow receiver
using nine-level sampling with two of the 2048 lag subcorrelators 
set to each polarization channel.
All observations were taken using the position-switching technique,
with the blank sky (or OFF) observation taken for the same length
of time, and over the same portion of the Arecibo dish (as defined
by the azimuth and zenith angles) as was used for the on-source (ON)
observation.  Each ON+OFF pair was followed by a 10s ON+OFF observation
of a well calibrated, uncorrelated noise diode.
The observing strategy used was as follows:
First, a minimum of one 3 minute ON/OFF pair was taken of each galaxy, 
followed by a 10s ON/OFF calibration pair.
If a galaxy was not detected, one or more additional 3 minute ON/OFF pairs
were taken of the object, if it was deemed of sufficient interest
(e.g., large diameter, known optical velocity).
If a galaxy's velocity was known a priori, it was observed
with each of the four 12.5 MHz bandpass subcorrelators being centred at the
redshifted \HI\ line of the galaxy.
All galaxies without known velocities were observed in velocity search mode.
In this mode the subcorrelators were set to 25MHz bandpasses.  Both
subcorrelators with the same polarization were set to overlap by only
5MHz, thereby allowing a wide velocity search while ensuring the overlapping
region of the two boards was adequately covered.
Two different velocity searched were made -- first in the
velocity range -500 to 11,000 \kms\ and subsequently in the range
9,500 to 21,000 \kms (assuming the galaxy was not detected in the
lower velocity range, the interest
of the particular object and observing time permitting).
The instrument's HPBW at 21 cm is \am{3}{6}$\times$\am{3}{6}
and the pointing accuracy is about 15$''$.

\subsection{Data reduction}  
Using standard data reduction software available at Arecibo,
the two polarizations were averaged, and corrections were applied
for the variations in the gain and system temperature of the telescope
with zenith angle and azimuth using the most recent calibration
data available at the telescope.
Further data analysis was performed using Supermongo routines developed by 
one of us (SES).
A baseline of order zero was fitted to the data,
excluding those velocity ranges with \HI\ line emission or
radio frequency interference (RFI).
Once the baselines were subtracted, the velocities were corrected to the
heliocentric system, using the optical convention, and the central line velocity,
line widths at, respectively, the 50\% and 20\% level of peak maximum (Lewis 1983), the
integrated flux of the \HI\ profiles, as well as the rms noise of the
spectra were determined. All data were boxcar smoothed to a velocity resolution
of 14.3 \kms\ (velocity search) and  16.4 \kms\ (known velocity) for analysis.

The stability of the chain of reception of the Arecibo telescope is known
to be good. This is shown by the observations we made of strong continuum
sources and of a calibration galaxy with a strong line signal: the latter showed a
$\pm$6\% standard deviation in its integrated line flux.

\section{Results}  
The global Arecibo \HI\ line data for the observed sample, together with their global 
near-infrared and optical data, are listed in Tables 3a,b for clearly detected objects, 
in Tables 4a,b for marginal detections and in Tables 5a,b for undetected 
objects.
A description of all parameters listed in the Tables is given below in Sections 
4.1-4.4. The near-infrared data listed were taken from the 2MASS catalogue
and the optical data were taken from the online NED and LEDA databases, as indicated.

The \HI\ spectra of the clearly detected galaxies are shown in Figure~2 and the marginal
detections are shown in Figure~3; for cases 
where 2 line profiles were detected in the same spectrum, designated by `a' and `b' after
the 2MASS sourcename, see Sect. 4.5.

The distribution of the integrated line fluxes and FWHM's of the clear and marginal detections 
is shown in Figure 4. Also plotted in this Figure is a straight line
indicating the 3 $\sigma$ detection limit for a 250 \kms\ wide, flat-topped spectral line
based on the average rms noise level of the data. The data quality and the rms noise 
of the Arecibo observations are rather uniform.

\subsection{Names, positions and distances}  
$\bullet$\, Number: we have divided the selected 2MASS sources according to 
two criteria: size and algorithm. This division is indicated
by two characters in the galaxy designations which we will use throughout this series: 
an entry starting with `L' indicates a `Large' object with an isophotal $K_s$-band radius
$r_{K_{20}}$ (see Sect. 4.2)  larger than 
20$''$, an `S' a `Small' one with a radius between 10$''$ and 20$''$ and an 'F' a galaxy 
selected using the LCSB source processor, while following the source number an `O'
 indicates  a previously catalogued object, a `P' one with a PGC entry only and an `N' 
a previously uncatalogued one.  

\noindent
$\bullet$\, Identifications: for each of the 2MASS sources we queried the NED 
and LEDA databases for a cross-identification of its position with other catalogues. 
For previously catalogued objects we list the most commonly used identification 
besides the 2MASS identification;

\noindent
$\bullet$\, Positions: the column headed `2MASSXi J' gives the 2MASS source designations, 
which are the right ascensions and declinations of their centre positions, for epoch J2000.0. 
These were used as the pointing centres for the \HI\ observations; 

\noindent
$\bullet$\, Distances: for each detected galaxy, a distance $D$ was calculated 
using radial \HI\ velocities corrected to the Galactic Standard of Rest, following the RC3 and 
assuming a Hubble constant of $H_0$=75~km~s$^{-1}$~Mpc$^{-1}$.

\subsection{Near-infrared data}  
\noindent
$\bullet$\, $K_{20}$ is the total magnitude measured within the $r_{K_{20}}$ isophotal 
aperture (see below);

\noindent
$\bullet$\, $J-K$ is a near-infrared colour of the source, based on magnitudes measured
in the $J$ and $K_s$ bands within their respective isophotal semi-major axes at the 
20 \masq\ level;


\noindent
$\bullet$\, $\mu_{K5}$ is the mean central surface brightness (in \masq) measured within a radius 
of 5 arcsec around the source's centre;

\noindent
$\bullet$\, $b/a$ is the infrared axis ratio determined from an ellipse fit to the co-addition of the 
$J$-, $H$-, and $K_s$-band images. The fit is carried out
at the 3-$\sigma$ isophotal level relative to the background noise in each
image. The 2MASS F (LCSB sources) sample was not measured in this way because of
the low signal-to-noise levels of the emission;

\noindent
$\bullet$\, $r_{K_{20}}$ is the fiducial aperture (in arcsec) in the $K_s$  band. Essentially, 
it is the aperture size for a surface brightness of 20 \masq;

\noindent
$\bullet$\, \LK\ is the absolute magnitude in the $K_s$band (in \LsunK), calculated
using an absolute solar $K_s$-band magnitude of 3.33 (Allen 1973).

\subsection{Optical data}  
$\bullet$\, $Type$ is the morphological type, as listed in NED or LEDA (the latter 
are between parenthesis)

\noindent
$\bullet$\, $V_{opt}$ is the mean heliocentric optical radial velocity (in \kms),
as listed in LEDA;

\noindent
$\bullet$\, $D_{25}$ is the isophotal $B$-band diameter (in units of arcmin) 
measured at a surface brightness level of 25 \masq, as listed in LEDA;

\noindent
$\bullet$\, $B_{T_{c}}$ is the total apparent $B$-band magnitude reduced to the
RC3 system (de Vaucouleurs et al. 1991) and corrected for galactic extinction, 
inclination and redshift effects (see Paturel et al. 1997, and references therein), 
as listed in LEDA;

\noindent
$\bullet$\, $\mu_{B_{25}}$ is the mean $B$-band surface brightness (in \masq)
within the 25 \masq\ isophote, as listed in LEDA;

\noindent
$\bullet$\, \LB\ is the absolute magnitude in the $B$ band (in \LsunB), calculated
using the $B_{T_{c}}$ magnitude and an absolute solar magnitude in the $B$ band of 5.48 
(Allen 1973), as listed in LEDA.

\subsection{\HIit\ data}  
The global \HI\ line parameters are directly measured values; 
no corrections have been applied to them for, e.g., instrumental 
resolution or cosmological stretching (e.g., Matthews et al. 2001).

\noindent
$\bullet$\, $rms$ is the rms noise level in a spectrum (in mJy);

\noindent
$\bullet$\, \IHI\ is the integrated line flux (in \Jykms).
The upper limits listed are 3$\sigma$ values for flat-topped profiles 
with a width of 250 \kms, a representative value for the galaxies detected;

\noindent
$\bullet$\, \VHI\ is the heliocentric central radial velocity of a line profile 
(in \kms), in the optical convention. We estimated the uncertainty, 
$\sigma_{V_{HI}}$  (in \kms), in \VHI\ following Fouqu\'e et al. (1990), as 
$\sigma_{V_{HI}} = 4 R^{0.5} P_{W}^{0.5} X^{-1}$, 
where $R$ is the instrumental resolution 
(16.4 \kms\ for galaxies with previously known velocity and 14.3 for velocity
searches ), 
$P_{W}$=($W_{20}$--$W_{50}$)/2 (in \kms) and $X$ is the signal-to-noise ratio of a 
spectrum, which we defined as the ratio of the peak flux density and the rms noise.

\noindent
$\bullet$\, $W_{50}$ and $W_{20}$ are the velocity widths (in \kms) at 
50\% and 20\% of peak maximum (km/s), respectively. According to Fouqu\'e et al. (1990),
the uncertainty in the linewidths is 2$\sigma_{V_{HI}}$ (see above) 
for W$_{50}$ and 3$\sigma_{V_{HI}}$ for W$_{20}$.
 
\noindent
$\bullet$\, \MHI\ is the total \HI\ mass (in \Msun), 
\MHI = 2.356 $10^5 D^2$ \IHI, where $D$ is in Mpc and \IHI\ in \Jykms;

\noindent
$\bullet$\, \MHILB\ is the ratio of the total \HI\ mass to the $B$-band 
luminosity (in \MsunLBsun);

\noindent
$\bullet$\, \MHILK\ is the ratio of the total \HI\ mass to the $K_s$-band 
luminosity (in \MsunLKsun);

\subsection{Notes on individual galaxies}  
In order to identify sources whose \HI\ detections could have been confused
by nearby galaxies,
we queried the NED and HyperLeda databases and inspected DSS images over a
region of 10$'$ radius surrounding the 2MASS position of each source.
Although the telescope's HPBW is \am{3}{6} only, an analysis we made of its beam 
pattern at the time of our observations shows that  in the first sidelobe, 
at 6$'$ radius, an estimated 20\% of the line flux would be picked up of a 1$'$ 
diameter galaxy, whereas at about \am{9}{5} radius this percentage has dropped to 0. 
We also queried these databases for published \HI\ line data of the target galaxies. 
For further information on galaxies with published \HI\ line parameters, see Section 5.2.
Quoted values are weighted averages from the HyperLeda database,
unless otherwise indicated.

{\sf L12O}:\, 2MASX J00234816+1441026 = UGC 226: UV excess galaxy Mrk 338 
(Mazzarella \& Balzano 1986), = IRAS F00212+1424.

{\sf L19O }:\, 2MASX J00323244+2323424 = UGC 321: 5 previous \HI\ detections at Arecibo (see Table 2). 
Optical H$\alpha$ rotation curve in Mathewson \& Ford (1996).

{\sf S298P:\, 2MASX J01512494+2235396 = PGC 212849: our \HI\ spectrum 
($V_{HI}$=9894 \kms, $I_{HI}$=2.3 \Jykms, $W_{50}$=348 \kms) is confused with that
of NGC 695, a $B_T$=13.9 mag, $D_{25}$=\am{0}{5} S0?pec galaxy with $V_{opt}$=9680$\pm$100 \kms\
at \am{2}{6} separation, for which average \HI\ parameters of 
$V_{HI}$=9737 \kms, $I_{HI}$=5.0 \Jykms\ and $W_{50}$=250 \kms\
are listed in HyperLeda, which are based on the Arecibo spectra of Garwood et al. (1987), 
Giovanelli et al. (1986), Lewis (1987) and Mirabel \& Sanders (1988).
We estimate that about 60\% of its line flux will be detected at the position of the target galaxy.}

{\sf S301O}:\, 2MASX J01545765+3720346 = UGC 1380: detected previously in \HI\ at Arecibo 
(Haynes et al. 1997). $H$-band photometry (Boselli et al. 2000;  Gavazzi et al. 2000).

{\sf L81O}:\, 2MASX J01553966+3702188 = CGCG 522-068: $H$-band photometry 
(Boselli et al. 2000;  Gavazzi et al. 2000). Not detected in our survey.

{\sf S490P}:\, 2MASX J03084665+2420483 = PGC 1705247: \HI\ profile partially outside our search range.

{\sf L224O}:\, 2MASX J04013761+2449189 = FGC464 / FGC 464b: our \HI\ spectrum of the target 
galaxy, NGC 464, shows two detections,  at 5803 and 6673 \kms, like the Arecibo spectrum of 
Giovanelli et al. (1997). At \am{0}{3} from the edge-on Sc galaxy lies another, somewhat smaller, 
edge-on spiral, FGC 464b. It is impossible to assign the detections to the objects, as neither has a
known optical redshift. For the distance dependent parameters listed for L224Oa and L224Ob
in Table 4b.1 it was assumed that the \HI\ is associated with NGC 464.

{\sf S662P}:\, 2MASX J04040187+1959363 = PGC 1612699: our \HI\ detection 
($V_{HI}$=6553 \kms, $I_{HI}$=1.4 \Jykms\ and $W_{50}$=230 \kms) is not expected to
be significantly confused with that of the S0 IC 358($V_{opt}$=6760$\pm$50 \kms), 
at \am{7}{4} separation, since only an estimated 8\% of its line flux will be detected 
at the position of the target galaxy. The Arecibo \HI\ profile of IC 358
(from Giovanelli \& Haynes 1993) shows
$V_{HI}$=6770 \kms, $I_{HI}$=3.6 \Jykms\ and $W_{50}$=638 \kms.

{\sf S686O}:\, 2MASX J04140133+2645003 = FGC 470: detected in \HI\ at Nan\c{c}ay 
(Matthews \& van Driel 2000).

{\sf L253P}:\, 2MASXI J0447309+242207 = PGC 1705745: \HI\ profile partially outside our search range.

{\sf S868P}:\, 2MASX J05221999+1749137 = PGC 1542586: at \am{2}{3}, i.e. 1.3 times half the 
telescope's HPBW from the target galaxy lies IRAS 05192+1745, an Sbc spiral with 
$V_{opt}$=5892$\pm$60 \kms, $B_T$=16.65 mag and $D_{25}$=\am{0}{8}.
As our \HI\ profile parameters are $V_{HI}$=5559 \kms\ and $W_{50}$=303 \kms, 
it does not appear to be confused with the nearby IRAS galaxy.

{\sf S766N}:\, 2MASX J04431049+2901446: our \HI\ detection 
($V_{HI}$=6505 \kms, $I_{HI}$=1.1 \Jykms, $W_{50}$=131 \kms) is not expected to
be significantly confused with that of the S0/a UGC 3142 ($V_{opt}$=6592$\pm$43 \kms), 
at \am{8}{7} separation, since only about 2.5\% of its line flux would be detected 
at the position of the target galaxy. The Arecibo \HI\ profiles of UGC 3142 
(from Lu et al. 1990 and Pantoja et al. 1997) show 
$V_{HI}$=6496 \kms, $I_{HI}$=5.1 \Jykms\ and $W_{50}$=288 \kms.

{\sf L313O \& S1017O}:\, 2MASX J07111558+3110140 \& 2MASX J07112782+3111218 = 
 CGCG 146-033 \& 146-036, respectively: 
the two target galaxies are members of an isolated triplet of galaxies, together with CGCG 146-034.
Our two \HI\ profiles do not appear to be confused. CGCG 146-034, at $V_{opt}$=7433$\pm$60 \kms, 
has been classified as an elliptical and is therefore likely to be devoid of \HI.
The \am{2}{9} separation between the two target galaxies is relatively large compared to the
telescope's half HPBW of \am{1}{8}, and their optical velocities (7515 and 7203 \kms, respectively)
are separated by 313 \kms. Our \HI\ profiles obtained pointing towards the centres of  
CGCG 146-033 and CGCG 146-036 in fact do not overlap in velocity: they have, respectively, 
$V_{HI}$=7496 and 7228 \kms, corresponding well to the optical systemic velocities, 
and $W_{50}$=163 and 197 \kms. The emission seen in spectrum of CGCG 146-036 at $\sim$7450-7600 \kms\ 
may be due to CGCG 146-033.
The two target 2MASS galaxies are included in the catalogue of nearby poor clusters of galaxies 
of White et al. (1999).
%

{\sf  S1017O}:\, 2MASX J07112782+3111218 = CGCG 148-26:
it is quite unlikely that our \HI\ detection 
($V_{HI}$=7181 \kms, $I_{HI}$=1.4 \Jykms\ and $W_{50}$=146 \kms) of the $B_T$=16.1 mag 
target galaxy, without know optical redshift, is confused by the somewhat brighter 
($B_T$$\sim$15.6) galaxies CGCG 148-33 and 148-34, at about 3$'$ distance, as their
optical redshifts of 7516$\pm$45 and 7433$\pm$60 \kms, respectively, are significantly higher.
No \HI\ data are available on these two objects.

{\sf S1025P}:\, 2MASX J07154721+1930583: it is not possible to determine if our \HI\ detection
($V_{HI}$=5207 \kms, $I_{HI}$=0.43 \Jykms\ and $W_{50}$=134 \kms) of the target galaxy
is confused with that of a galaxy of similar optical size and brightness,
2MASX J07160231+1932242, at \am{3}{8} distance, since neither has a known optical redshift.
At this separation, we estimate that about 30\% of  line flux of the latter object would be detected 
at the target position.

{\sf S1062O}:\, 2MASX J07471628+2657037 = IC 746: our \HI\ detection appears to be of nearby spiral 
galaxy NGC 2449, rather than of the target 2MASS object. Our spectrum centered on the 2MASS galaxy 
shows $V_{HI}$=4776 \kms, $I_{HI}$=1.4 \Jykms\ and $W_{50}$=248 \kms, while its optical redshift
is 229 \kms\ higher, 5005$\pm$39 \kms. NGC 2449, which lies only \am{1}{5} from the target 
galaxy and well within the Arecibo HPBW, is an Sab spiral of $B_T$=14.13 mag and $D_{25}$=\am{1}{4}, with 
$V_{opt}$=4817$\pm$34 \kms. A detection of NGC 2449 at Arecibo (Bicay \& Giovanelli 1986),
shows $V_{HI}$=4778 \kms, $I_{HI}$=2.0 \Jykms\ and $W_{50}$=221 \kms.
Two other galaxies of similar redshift, NGC 2449 ($V_{opt}$= 4801$\pm$33 \kms) and 
IC 2205 ($V_{opt}$=4726$\pm$43 \kms), both about one magnitude fainter in $B$ than NGC 2449,
lie about 6$'$ from the target galaxy. No \HI\ observations are published of these objects.
The target galaxy is included in the catalogue of nearby poor clusters of galaxies 
of White et al. (1999).

{\sf L324O}:\,  2MASX J07480326+2801422 = CGCG 148-26:
our \HI\ profile ($V_{HI}$=8041 \kms, $I_{HI}$=1.7 \Jykms\ and $W_{50}$=153 \kms)
of the target galaxy ($V_{opt}$=8111$\pm$60 \kms, $B_T$=16.4) may in principle be confused 
by a nearby (\am{3}{3} separation), somewhat more luminous galaxy galaxy, CGCG 148-28
($V_{opt}$=8163$\pm$43 \kms, $B_T$=15.6), both without published morphological type.
At this separation, we expect that about 40\% of the line flux of the latter would be detected 
at the position of the target galaxy.

%

{\sf S1080O}:\, 2MASX J07551046+1422131 = CGCG 58-70: 
there will be no confusion of our detection at $V_{HI}$=13,446 \kms\ of
the target galaxy ($V_{opt}$=13,447 \kms) with the three similarly bright galaxies at about 4$'$ to
\am{7}{5} distance (CGCG 58-68, -69 and -71), since these all have optical redshifts of 
$\sim$8750 \kms.

{\sf L363O}:\, 2MASX J09030837+1739356= FGC 828:
since none have optical redshifts, it is impossible to determine if our \HI\ detection
($V_{HI}$=8832 \kms, $I_{HI}$=1.0 \Jykms\ and $W_{50}$=158 \kms) of the target galaxy,
with $K_{20}$=12.5, is confused by 2 smaller nearby 2MASS galaxies, 2MASX J09030383+1741336
of $K_{20}$=12.8 at \am{2}{3} distance, and 2MASX J09031111+1740431
of $K_{20}$=14.4 at \am{1}{4} distance. At these separations, we estimate that about 
70\% to 90/
The galaxy pair CGCG 90-59, at \am{6}{5} distance, has a much higher redshift of about
16,000 \kms.

{\sf S1283O}:\, 2MASX J09223543+2051068 = PGC 86616: detected previously in \HI\ at Arecibo 
(Schombert et al. 1992).

{\sf S1317O}:\, 2MASXI J0939167+321838 = NGC 2944 pair: interacting galaxy pair (= Arp 63).
The separation of NGC 2944 into two galaxies is best seen on the 2MASS images, where the 
projected distance of their nuclei is 16$''$, as the DSS image is saturated in the inner 
regions. The selected 2MASS source corresponds to the centre of the least bright, W nucleus.
Optical spectroscopy (Maehara et al. 1987) only showed H$\alpha$+$[$N{\sc ii}$]$ emission 
lines, with an equivalent width of 57\AA\ for H$\alpha$.
28$''$ SE of the centre of the NGC 2944 pair lies PGC 27534, a $B_T$=16: object of
$D_{25}$=\am{0}{2} without published optical redshift.
NGC 2944 was detected previously in \HI\ at Arecibo (Williams \& Kerr 1981) and at 
Nan\c{c}ay, (Chamaraux 1977). The central \HI\ velocities measured at the two 
telescope differ significantly: 
6783 and 6753 \kms\ at Arecibo compared to 6831 \kms\ at Nan\c{c}ay. 
The much higher Nan\c{c}ay $I_{HI}$ of 13.6 \Jykms\ compared to the 4.0 \Jykms\
we measured at Arecibo can be explained partially by confusion within the larger 
Nan\c{c}ay beam with the interacting UGC 5146 pair (=Arp 129), at \am{3}{7} separation, 
for which Bicay \& Giovanelli (1986) measured at Arecibo $V_{HI}$=6862, 
$I_{HI}$=4.8 \Jykms\ and $W_{50}$=163 \kms.
The $W_{50}$ width we measured at Arecibo is almost twice the value measured by
Williams \& Kerr, 149 /kms.

{\sf L388O}:\, 2MASX J09403295+2529255 = UGC 5156: its optical redshift is 
10,028$\pm$43 \kms. It was detected previously in \HI\ at Arecibo 
(Bicay \& Giovanelli 1986), at $V_{HI}$=9953 \kms, 174 \kms\ higher than our
central velocity, which appears to be due to the fact that our \HI\ profile lies partially 
outside the search range.
At \am{5}{8} distance lies a galaxy of similar redshift ($V_{opt}$=10,082$\pm$43 \kms), 
size, magnitude and high inclination, CGCG 122-47. The \HI\ profiles may in principle be
confused by this object, as we estimate that about 25\% of its  line flux would be 
detected at the target position.

{\sf S1342O}:\, 2MASX J09501016+3424534 =  KUG 0947+346: its optical velocity, 6590$\pm$47 \kms\
(Kowal et al. 1974) is about 1000 \kms\ lower than our \HI\ velocity, 
7571 \kms. At \am{7}{1} distance lies a somewhat brighter Sc spiral, KUG 0946+346, of unknown
redshift. Since we estimate that only about 15\% of its line flux would be 
detected at the target position, it does not seem likely that the \HI\ profile is
confused by it.

{\sf S1357O}:\, 2MASX J09560425+1630537 = LSBC F637-01: our \HI\ detection 
($V_{HI}$=3825 \kms, $I_{HI}$=1.2 \Jykms\ and $W_{50}$=130 \kms) is not expected to
be significantly confused by the SBa NGC 3053 ($V_{opt}$=3774$\pm$60 \kms), 
at \am{8}{8} separation, since only an estimated 2.5\% of its line flux will be detected 
at the target position. The Arecibo \HI\ profiles of NGC 3053
(Chengalur et al. 1993 and Giovanardi \& Salpeter 1985) show 
$V_{HI}$=3730 \kms, $I_{HI}$=4.3 \Jykms\ and $W_{50}$=409 \kms.

{\sf S2687P}:\, 2MASX J22401444+3438401 = PGC 2052363:
our \HI\ profile ($V_{HI}$=8404 \kms, $I_{HI}$=2.6 \Jykms, $W_{50}$=226 \kms)
may in principle be confused by the galaxy 2MASX J22404656+3437520 at \am{6}{4}
distance, which is 1.2 mag brighter in the $K_s$ band.  Neither have an optical
redshift. Only about an estimated 15\% of its line flux would be 
detected at the target position.

{\sf S2730P}:\,  2MASX J22551306+3248080 = PGC 2012878: our \HI\ profile may be a detection
of nearby galaxy Ark 569, an SBbc spiral with $B_T$=15.6 and $D_{25}$=\am{0}{8}
without a known optical redshift. At \am{4}{3} distance, we estimate that about 30\% of
its line flux would be detected at the target position.
An Arecibo detection (Giovanelli \& Haynes 1985) of Ark 569 shows $V_{HI}$=6000 \kms\ 
and $W_{50}$=266 \kms, values very close to the $V_{HI}$=6018 \kms\ 
and $W_{50}$=270 \kms\ we measured pointing towards the 2MASS object.
The integrated \HI\ line flux we measured towards the 2MASS galaxy, 0.4 \Jykms,
is 10 times smaller than the 4.0 \Jykms\ measured towards Ark 569.

{\sf L802P}:\, 2MASX J22580271+3227586 = PGC 2800849: detected previously in \HI\ at Arecibo 
(Cabanela \& Dickey 1999), with $V_{HI}$=6588 \kms\ and $I_{HI}$=1.5 \Jykms, and an 
average of the $W_{50}$ and $W_{20}$ line widths of 283 \kms.
Surprisingly, the object was not detected in our survey, with an rms noise level of 1.19 mJy.
For a flat-topped profile with width of 283 \kms, this would imply a 3$\sigma$ upper
limit to $I_{HI}$ of 1.0 \Jykms, well below the value measured previously.

{\sf L807O}:\,  2MASX J22594147+2404297 = UGC 12289: our \HI\ profile of the
target galaxy ($V_{HI}$=9986 \kms, $W_{50}$=217 \kms, $I_{HI}$=4.3 \Jykms), with 
$V_{opt}$=10,160$\pm$42 \kms, lies partially outside 
our search range, which explains the 174 \kms\ difference with the
optical redshift and the Arecibo detection by Giovanelli et al. (1986), which shows 
$V_{HI}$=10,160 \kms, $W_{50}$=218 \kms\ and $I_{HI}$=3.1 \Jykms. 
The profile is not expected to be confused significantly by the $B_T$ 16.1 SBab spiral
UGC 12283, with $V_{opt}$=9949$\pm$60 \kms, which was detected at Arecibo by
Giovanelli et al. (1986), who found $V_{HI}$=9949 \kms, $W_{50}$=274 \kms\ and 
$I_{HI}$=0.7 \Jykms. At \am{5}{3} distance, we estimate that about 25\% of
its line flux would be detected at the target position.

{\sf L815O}:\,  2MASX J23145953+1459192 = MCG 02-59-12: our \HI\ detection 
($V_{HI}$=11,908 \kms, $I_{HI}$=0.95 \Jykms, $W_{50}$=48 \kms) of the target galaxy,
which has an optical redshift of 12,182$\pm$60 \kms, may in principle
be confused by two nearby objects: a $B$=16 mag object without a PGC entry
at 23$^h$15$^m$\tis{12}{2}, 14$^{\circ}$58$'$22$''$ at 12,182 \kms\ and \am{3}{2}
distance, and $B_T$=16.2 mag Sc spiral UGC 12448 at 11,955 \kms\ and \am{6}{4}
distance. At these separations, we estimate that about 40\% and 20\%, respectively, 
of  their line fluxes would be detected at the target position.

{\sf S2817O}:\, 2MASXI J2323068+124955 =  KUG 2320+125: $B$ and $R$-band CCD imaging and 
long-slit spectroscopy (Grogin \& Geller 1999, 2000).

{\sf L829O}:\,  2MASXI J2347570+280747 = FGC 2536: detected in \HI\ at Nan\c{c}ay 
(Matthews \& van Driel 2000).

{\sf S2834O}:\, 2MASX J23343591+2353362 = KUG 2332+236: two \HI\ lines were detected
in our spectrum, at 5615 and 9802 \kms, respectively. The $B_T$=16.5 Sc-type target galaxy 
has no known optical redshift. For the distance dependent parameters listed for S2834Oa and b
in Table 4b.2 it was assumed that the \HI\ is associated with the target galaxy.
Our lower velocity detection ($V_{HI}$=5615 \kms, $I_{HI}$=0.9 \Jykms, $W_{50}$=298 \kms)
will be completely confused by two nearby galaxies, both without known optical redshift, which 
were detected in \HI\ at Arecibo by Spitzak \& Schneider (1998): the $B_T$=17.3 mag PGC 169947  
($V_{HI}$=5603 \kms, $I_{HI}$=1.7 \Jykms, $W_{50}$=162 \kms) at \am{2}{1},
and the $B_T$=17.6 mag PGC 169948  
($V_{HI}$=5558 \kms, $I_{HI}$=2.7 \Jykms, $W_{50}$=109 \kms) at \am{5}{1} distance.
We estimate that about 75\% of the line flux of the former and about 25\% of the flux of
the latter would be detected at the target position, i.e. in total about 2 \Jykms - twice the
total flux we measured.

\section{Discussion}  

\subsection{Comparison with published \HIit\ data}  
Of the 127 galaxies we detected at 21 cm, 10 were detected 
previously in \HI\ (see Sect. 4.5 and Table 2).
Of these, 7 were detected at Arecibo only, one at both Arecibo and Nan\c{c}ay 
(NGC 2944) and 2 at Nan\c{c}ay only (FGC 470 and 2536). 
Basically all published Arecibo measurements were made
before its major upgrade, while ours were made after. A comparison of these data with
ours generally shows good agreement, except for the case of the confused \nan\ profile
of the interacting pair NGC 2944 (S1317O) and for UGC 12289 (L807O), where our 
profile lies partially outside the band (see Section 4.5). 
Excluding NGC 2944 and UGC 12289, the average absolute value and its $\sigma_N$ 
deviation of the differences between our radial velocities and the published 
\HI\ velocities is 25$\pm$21 \kms, compared to the average estimated uncertainty 
of 6 \kms\ in the values we measured for these objects, while the average ratio 
of our $W_{50}$ line widths and the published values is 1.02$\pm$0.06, and the 
ratio of the integrated line fluxes is 0.82$\pm$0.17.

\subsection{Radio Frequency Interference (RFI)} 
As a consequence of their high sensitivity, radio astronomy telescopes are 
vulnerable to radio frequency interference (RFI), with signal strengths usually greatly 
exceeding those of the weak celestial radio sources being observed. Broad-band RFI raises 
the noise level of the observations, while narrow-band RFI may mimic spectral lines
like the \HI\ lines from galaxies that are being searched for in the present study.

It should be noted that in the ITU-R Radio Regulations (2001), with which
all users of the radio spectrum are obliged to comply, astronomical \HI\ line 
observations are only protected from ``all emissions'' up to a redshift of about 
4300 \kms, while for observations further out to 
$\sim$19,000 \kms\ ``[national frequency] administrations are urged to take 
all practicable steps to protect the radio astronomy service from harmful 
interference''. These regulatory provisions for the protection of the Radio 
Astronomy Service clearly cannot guarantee a completely RFI-free environment for
the kind of survey we performed.

Though care was taken to make the renovated Arecibo telescope more robust
for unwanted radio interference (RFI), and to coordinate its operation as 
well as possible with the frequency plan and emission periods of local radar 
installations, persistent RFI signals with strengths that make the detection 
of faint \HI\ line signals impossible were present during a significant 
fraction of the observations. 

In order to examine during what percentage of time which radial velocity ranges 
are interfered with, considering the faintness of the line signals we are 
searching for, we flagged all channels in all spectra made in the velocity 
search mode (see Sect. 3.1) showing signals with a flux density 
exceeding the 20 mJy level after baseline fitting -- a level exceeded by 
only two of our 107 detected galaxies: S622P at 11,711 \kms\ and L807O at 9986 \kms.
Plots of the relative number of times such 
signals occurred during the November 2000 and January 2001 observing runs are shown 
in Figure 1, in bins with a a width of 25 \kms\ (about 10 channels).
Besides RFI, the plots show the omnipresent Galactic \HI\ emission around 0 \kms.
The most troublesome RFI signal in the velocity searches, of which about 85\% were 
made in the -500 to 11,000 \kms\ range, are the intermittent GPS L3 emissions in the 
$\sim$81500-8700 \kms\ range (around 1381 MHz): 12\% of all spectra showed RFI 
exceeding the 20 mJy level in this range.


\section{Conclusions}
Of the 367 galaxies observed, 127 (35\%) were clearly detected,
66 of which did not have a previously known radial velocity.
The detection statistics as function of type (previously catalogued , PGC entry only or
uncatalogued) and size or selection algorithm (large/small/faint) are listed in Table 1. 
For the Large objects ($r_{K_{20}}\geq 20''$) the global detection rate 
(for both previously catalogued and uncatalogued objects) is  79\%,
while for the small objects ($20''\geq r_{K_{20fe}} \geq 10''$), the global 
detection rate is much lower, 26\%.

\acknowledgements{ 
This publication makes use of data products from the Two Micron 
All Sky Survey, which is a joint project of the University of Massachusetts 
and the Infrared Processing and Analysis Center, funded by the National 
Aeronautics and Space Administration and the National Science Foundation.
We also wish to thank the Arecibo Observatory which is part of the National 
Astronomy and Ionosphere Center, which is operated by Cornell University under 
a cooperative agreement with the National Science Foundation.  
This research also has made use of the Lyon-Meudon Extragalactic 
Database (LEDA), recently incorporated in HyperLeda,  the NASA/IPAC 
Extragalactic Database (NED)   
which is operated by the Jet Propulsion Laboratory, California Institute   
of Technology, under contract with the National Aeronautics and Space      
Administration and the Aladin database, operated at CDS, Strasbourg, France.  
We acknowledge financial support from CNRS/NSF collaboration grant No.
10637 and from the ASTE of CNRS/INSU.
}

\newpage

\onecolumn

\begin{table}
\centering
\bigskip
{\footnotesize
\begin{tabular}{lrrrrrrrrr}
\multicolumn{10}{l}{{\bf Table 1.} Detection statistics}\\
\smallskip \\
\hline
\vspace{-2 mm} \\
 & \multicolumn{3}{c} {Previously} & \multicolumn{3}{c} {Previously} & \multicolumn{3}{c} {Previously}\\ 
 & \multicolumn{3}{c} {Catalogued} & \multicolumn{3}{c} {Catalogued} & \multicolumn{3}{c} {Uncatalogued}\\
 & \multicolumn{3}{c} {(Other)} & \multicolumn{3}{c} {(PGC)}& \multicolumn{3}{c} {(New)}\\  
 & \multicolumn{1}{c} {Large} & \multicolumn{1}{c} {Small} & \multicolumn{1}{c} {Faint} 
 & \multicolumn{1}{c} {Large} & \multicolumn{1}{c} {Small} & \multicolumn{1}{c} {Faint}   
 & \multicolumn{1}{c} {Large} & \multicolumn{1}{c} {Small} & \multicolumn{1}{c} {Faint} \\   
\vspace{-2 mm} \\
 \hline
\vspace{-2 mm} \\
 Total observed & 25 & 75 & 1 & 14 & 179 & 0 & 17 & 55 & 1\\
\vspace{-2 mm} \\
\hline
\vspace{-2 mm} \\
 Detected    & 19 & 37 & 1 & 8 & 27 & 0 & 12 & 2 & 1 \\
 Marginal    & 2 & 7 & 0 & 0 & 8 & 0 & 3 & 0 & 0 \\
 Undetected  & 4 & 31 & 0 & 6 & 144 & 0 & 2 & 53 & 0 \\
\vspace{-2 mm} \\
 \hline
\vspace{-2 mm} \\
 Det.+Margin. & 21 & 44 & 1 & 8 & 35 & 0 & 15 & 2 & 1 \\
             & $84\%$ & $59\%$ & $100\%$ & $57\%$ & $20\%$ & --- & $88\%$ & $4\%$ & $100\%$\\
\vspace{-2 mm} \\
\hline
\vspace{-2 mm} \\
\multicolumn{ 10}{l}{{\bf Note:} Large : $r_{K_{20}}\geq 20''$, Small: $20''\geq r_{K_{20}}\geq
10''$, Faint: galaxies selected with the } \\
\multicolumn{ 10}{l}{LCSB source processor.} \\
\end{tabular}
}
\normalsize
\end{table} 

\newpage
\clearpage

\begin{table}
\centering
\bigskip
{\scriptsize
\begin{tabular}{llrlrll}
\multicolumn{7}{l}{\footnotesize {\bf Table 2.} Comparison with published 
 \HI\ data}\\
\smallskip \\
\hline
\vspace{-2 mm} \\     
No. & Ident. & $V_{HI}$ & $I_{HI}$ & $W_{50}$ & Tel. & Ref. \\
    &   & km/s    & \JYKMS\ &  km/s & &   \\
\vspace{-2 mm} \\
\hline 
\vspace{-2 mm} \\
L19O & UGC 321 & 4655 & 2.6 & 219 & A & this paper  \\
\end{tabular}
}
\end{table}

%
\begin{table}
\centering
\bigskip
{\scriptsize
\begin{tabular}{llrl}
\multicolumn{4}{l}{\footnotesize {\bf Table 3a.1.} Detected objects } \\
\multicolumn{4}{l}{\footnotesize   \,\,\,\,\,\,\,\,\,\,\,\,\,\,\,\,\,\,\,\,\,\,\,\,\,\,\,\, 
                                  -- identifications and coordinates}\\
\smallskip \\
\hline
\vspace{-2 mm} \\
Number & 2MASS & PGC & others  \\ 
& hhmmss.s+ddmmss &  &  \\  
 & (J2000.0) &  &  \\ 
\vspace{-2 mm} \\
\end{tabular}
}
\normalsize
\end{table}

%
\begin{table}
\centering
\bigskip
{\scriptsize
\begin{tabular}{llrl}
\multicolumn{4}{l}{\footnotesize {\bf Table 3a.2.} Detected objects}\\
\multicolumn{4}{l}{\footnotesize \,\,\,\,\,\,\,\,\,\,\,\,\,\,\,\,\,\,\,\,\,\,  
                                      -- identifications and coordinates - {\it continued} }\\
\smallskip \\
\hline
\vspace{-2 mm} \\
Number & 2MASS & PGC & others  \\ 
& hhmmss.s+ddmmss &  &  \\  
 & (J2000.0) &  &  \\ 
\vspace{-2 mm} \\
\end{tabular}
}
\normalsize
\end{table} 
 
%
\begin{sidewaystable}
\centering
\bigskip
{\scriptsize
\begin{tabular}{lrrrrrrrrlrrrrrrrrrrrrr}
\multicolumn{23}{l}{\footnotesize {\bf Table 3b.1.} Detected objects -- basic data}\\
\smallskip \\
\hline
\vspace{-2 mm} \\
No & $K_{20}$ & $J-K$ & $\mu_{K5}$ & $b/a$ & $r_{K_{20}}$ & $B_{T_{c}}$ & $D_{25}$ & $\mu_{B_{25}}$ & $T$ & rms & 
 $I_{HI}$ & $V_{HI}$ & $\sigma_{V_{HI}}$ & $V_{opt}$ & $W_{50}$ & $W_{20}$ & $D$ & \MHI\ & \LK\ & \LB\ & \MHILKfr\ & \MHILBfr\  \\
 & & & & & & & & & & & & & & & & & & [log] & [log] & [log] & & \\
  & mag & mag & $\frac{mag}{''^2}$ & & $''$ & mag & $'$ & $\frac{mag}{''^2}$ & & mJy & 
 \JYKMS\ & \KMS\ & \KMS\ & \KMS\ & \KMS\ & \KMS\ & Mpc & \Msun\ & \LsunK\ & \LsunB\ & & \\ 
\vspace{-2 mm} \\
\hline
\vspace{-2 mm} \\
\end{tabular}
}
\normalsize
\end{sidewaystable}

\begin{sidewaystable}
\centering
\bigskip
{\scriptsize
\begin{tabular}{lrrrrrrrrlrrrrrrrrrrrrr}
\multicolumn{23}{l}{\footnotesize {\bf Table 3b.2.} Detected objects -- basic data - {\it continued} }\\
\smallskip \\
\hline
\vspace{-2 mm} \\
No & $K_{20}$ & $J-K$ & $\mu_{K5}$ & $b/a$ & $r_{K_{20}}$ & $B_{T_{c}}$ & $D_{25}$ & $\mu_{B_{25}}$ & $T$ & rms & 
 $I_{HI}$ & $V_{HI}$ & $\sigma_{V_{HI}}$ & $V_{opt}$ & $W_{50}$ & $W_{20}$ & $D$ & \MHI\ & \LK\ & \LB\ & \MHILKfr\ & \MHILBfr\  \\
 & & & & & & & & & & & & & & & & & & [log] & [log] & [log] & & \\
  & mag & mag & $\frac{mag}{''^2}$ & & $''$ & mag & $'$ & $\frac{mag}{''^2}$ & & mJy & 
 \JYKMS\ & \KMS\ & \KMS\ & \KMS\ & \KMS\ & \KMS\ & Mpc & \Msun\ & \LsunK\ & \LsunB\ & & \\ 
\vspace{-2 mm} \\
\hline
\vspace{-2 mm} \\
\end{tabular}
}
\normalsize
\end{sidewaystable}

%
\begin{table}
\centering
\bigskip
{\scriptsize
\begin{tabular}{llrl}
\multicolumn{4}{l}{{\bf Table 4a.} Marginal detections -- 
                                   identifications and coordinates}\\
\smallskip \\
\hline
\vspace{-2 mm} \\
Number & 2MASS & PGC & others  \\ 
& hhmmss.s+ddmmss &  &  \\  
 & (J2000.0) &  &  \\ 
\vspace{-2 mm} \\
\hline
\vspace{-2 mm} \\
S137P  &  0055337+294438 &   1876630 &               \\
S370N  &  0219284+164731 &           &               \\
S388O  &  0225429+284034 &      9223 & KUG 0222+284  \\
S446P  &  0248590+151241 &   1478344 &               \\
S658P  &  0402598+134515 &   1439552 &               \\
S676N  &  0409533+243427 &           &               \\
S1035O &  0721585+292950 &     20814 & CGCG 147-013  \\
S1086P &  0758032+144144 &   1464653 &               \\
S1100O &  0805478+301350 &     22726 & CGCG 148-111  \\
S1206O &  0855023+264031 &     25028 & CGCG 150-045  \\
S1243O &  0907047+360646 &     86893 & KUG 0903+363B \\
S1259P &  0915232+312107 &   1943146 &               \\
L383O  &  0929352+243525 &     26944 & CGCG 122-004  \\
S1303O &  0932452+333820 &     82424 & KUG 0929+338  \\
S1311O &  0936581+311952 &     27367 & KUG 0934+315A \\
L386O  &  0938557+354141 &     82492 & KUG 0935+359  \\
S2730P &  2255130+324808 &   2012878 &               \\
L822O  &  2325102+121820 &     71383 & CGCG 431-060  \\
S2823P &  2328052+173440 &   1536138 &               \\
S2851P &  2340459+263318 &     93864 &               \\
S2877N &  2352344+163651 &           &               \\
\vspace{-2 mm} \\
\hline
\vspace{-2 mm} \\
\end{tabular}
}
\normalsize
\end{table}

%
\begin{sidewaystable}
\centering
\bigskip
{\scriptsize
\begin{tabular}{lrrrrrrrrlrrrrrrrrrrr}
\multicolumn{21}{l}{\footnotesize {\bf Table 4b.} Marginal detections -- basic data}\\
\smallskip \\
\hline
\vspace{-2 mm} \\
No & $K_{20}$ & $J-K$ & $\mu_{K5}$ & $b/a$ & $r_{K_{20}}$ & $B_{T_{c}}$ & $D_{25}$ & $\mu_{B_{25}}$ & $T$ & rms & 
 $I_{HI}$ & $V_{HI}$ & $V_{opt}$ & $W_{50}$ & $D$ & \MHI\ & \LK\ & \LB\ & \MHILKfr\ & \MHILBfr\  \\
 & & & & & & & & & & & & & & & & [log] & [log] & [log] & & \\
  & mag & mag & $\frac{mag}{''^2}$ & & $''$ & mag & $'$ & $\frac{mag}{''^2}$ & & mJy & 
 \JYKMS\ & \KMS\ & \KMS\ & \KMS\ & Mpc & \Msun\ & \LsunK\ & \LsunB\ & & \\
\vspace{-2 mm} \\
\hline
\vspace{-2 mm} \\
\end{tabular}
}

\normalsize
\end{sidewaystable}

\begin{table}                                                                                                                                                                                           
\centering                                                                                                                                                                                              
\bigskip                                                                                                                                                                                                
{\scriptsize                                                                                                                                                                                            
\begin{tabular}{llrl}                                                                                                                                                                                  
\multicolumn{4}{l}{\footnotesize {\bf Table 5a.1.} Undetected objects}\\                                                                                                                       
\multicolumn{4}{l}{\footnotesize  \,\,\,\,\,\,\,\,\,\,\,\,\,\,\,\,\,      
                                  -- identifications and coordinates}\\                                                                                                                                
\smallskip \\                                                                                                                                                                                           
\hline                                                                                                                                                                                                  
\vspace{-2 mm} \\                                                                                                                                                                                       
Number & 2MASS & PGC & others  \\                                                                                                                                                        
 & hhmmss.s+ddmmss &  &  \\                                                                                                                                                                      
 & (J2000.0) &  &  \\                                                                                                                                                                      
\vspace{-2 mm} \\
\end{tabular}                                                                                                                                                                                           
}                                                                                                                                                                                                       
\normalsize                                                                                                                                                                                             
\end{table}                                                                                                                                                                                             
                
%
\begin{table}                                                                                                                                                                                           
\centering                                                                                                                                                                                              
\bigskip                                                                                                                                                                                                
{\scriptsize                                                                                                                                                                                            
\begin{tabular}{llrl}                                                                                                                                                                                  
\multicolumn{4}{l}{\footnotesize {\bf Table 5a.2.} Undetected objects}\\                                                                                                                               
\multicolumn{4}{l}{\footnotesize \,\,\,\,\,\,\,\,\,\,\,\,\,\,\,\,\, 
                                   -- identifications and coordinates - {\it continued} }\\                                                                                                                            
\smallskip \\                                                                                                                                                                                           
\hline                                                                                                                                                                                                  
\vspace{-2 mm} \\                                                                                                                                                                                       
Number & 2MASS & PGC & others  \\ 
& hhmmss.s+ddmmss &  &  \\  
 & (J2000.0) &  &  \\ 
\vspace{-2 mm} \\
\end{tabular}                                                                                                                                                                                           
}                                                                                                                                                                                                       
\normalsize                                                                                                                                                                                             
\end{table}                                                                                                                                                                                             
 
%
\begin{table}                                                                                                                                                                                           
\centering                                                                                                                                                                                              
\bigskip                                                                                                                                                                                                
{\scriptsize                                                                                                                                                                                            
\begin{tabular}{llrl}                                                                                                                                                                                  
\multicolumn{4}{l}{\footnotesize {\bf Table 5a.3.} Undetected objects }\\                                                                                                                               
\multicolumn{4}{l}{\footnotesize \,\,\,\,\,\,\,\,\,\,\,\,\,\,\,\,\, -- identifications and coordinates 
                     - {\it continued} }\\                                                                                                                            
\smallskip \\                                                                                                                                                                                           
\hline                                                                                                                                                                                                  
\vspace{-2 mm} \\                                                                                                                                                                                       
Number & 2MASS & PGC & others  \\ 
& hhmmss.s+ddmmss &  &  \\  
 & (J2000.0) &  &  \\ 
\vspace{-2 mm} \\
\end{tabular}                                                                                                                                                                                           
}                                                                                                                                                                                                       
\normalsize                                                                                                                                                                                             
\end{table}                                                                                                                                                                                             
                                                                                                                                                                                                        
%
\begin{table}                                                                                                                                                                                           
\centering                                                                                                                                                                                              
\bigskip                                                                                                                                                                                                
{\scriptsize                                                                                                                                                                                            
\begin{tabular}{llrl}                                                                                                                                                                                  
\multicolumn{4}{l}{\footnotesize {\bf Table 5a.4.} Undetected objects }\\                                                                                                                               
\multicolumn{4}{l}{\footnotesize \,\,\,\,\,\,\,\,\,\,\,\,\,\,\,\,\, -- identifications and coordinates 
                     - {\it continued} }\\                                                                                                                            
\smallskip \\                                                                                                                                                                                           
\hline                                                                                                                                                                                                  
\vspace{-2 mm} \\                                                                                                                                                                                       
Number & 2MASS & PGC & others  \\ 
& hhmmss.s+ddmmss &  &  \\  
 & (J2000.0) &  &  \\ 
\vspace{-2 mm} \\
\end{tabular}                                                                                                                                                                                           
}                                                                                                                                                                                                       
\normalsize                                                                                                                                                                                             
\end{table}

%
\begin{sidewaystable}                                                                                                                                                                                   
\centering                                                                                                                                                                                              
\bigskip                                                                                                                                                                                                
{\scriptsize                                                                                                                                                                                            
\begin{tabular}{lrrrrrrrrrrccccl}                                                                                                                                                                       
\multicolumn{16}{l}{\footnotesize {\bf Table 5b.1.} Undetected objects -- basic data}\\                                                                                                           
\smallskip \\                                                                                                                                                                                           
\hline                                                                                                                                                                                                  
\vspace{-2 mm} \\                                                                                                                                                                                       
No & $K_{20}$ & $J-K$ & $H-K$ & $\mu_{K5}$ & $b/a$ & $r_{K_{20}}$ & $B_{T_{c}}$ & $D_{25}$ & $\mu_{B_{25}}$ & $T$ & rms &                                                                                          
 $I_{HI}$ &    \MHILKfr\  &   \MHILBfr\  & search\\                                                                                                                                                         
    & mag & mag & mag & $\frac{mag}{''^2}$  &  & $''$ & mag & $'$ & $\frac{mag}{''^2}$ &   & mJy & Jy km/s &   &  & mode \\                                                                                                                                                                                   
\vspace{-2 mm} \\
\hline                                                                                                                                                                                                  
\end{tabular}                                                                                                                                                                                           
}                                                                                                                                                                                                       
\normalsize                                                                                                                                                                                             
\end{sidewaystable}

%
\begin{sidewaystable}                                                                                                                                                                                   
\centering                                                                                                                                                                                              
\bigskip                                                                                                                                                                                                
{\scriptsize                                                                                                                                                                                            
\begin{tabular}{lrrrrrrrrrrccccl}                                                                                                                                                                       
\multicolumn{16}{l}{\footnotesize {\bf Table 5b.2.} Undetected objects 
                    -- basic data - {\it continued}}\\                                                                                         
\smallskip \\                                                                                                                                                                                           
\hline                                                                                                                                                                                                  
\vspace{-2 mm} \\                                                                                                                                                                                       
No & $K_{20}$ & $J-K$ & $H-K$ & $\mu_{K5}$ & $b/a$ & $r_{K_{20}}$ & $B_{T_{c}}$ & $D_{25}$ & $\mu_{B_{25}}$ & $T$ & rms &                                                                                          
 $I_{HI}$ &    \MHILKfr\  &   \MHILBfr\  & search \\                                                                                                                                                         
    & mag & mag & mag & $\frac{mag}{''^2}$  &  & $''$ & mag & $'$ & $\frac{mag}{''^2}$ &   & mJy & Jy km/s   &   &  & mode \\                                                                                                                                                                                   \vspace{-2 mm} \\
\end{tabular}                                                                                                                                                                                           
}                                                                                                                                                                                                       
\normalsize                                                                                                                                                                                             
\end{sidewaystable}

%
\begin{sidewaystable}                                                                                                                                                                                   
\centering                                                                                                                                                                                              
\bigskip                                                                                                                                                                                                
{\scriptsize                                                                                                                                                                                            
\begin{tabular}{lrrrrrrrrrrccccl}                                                                                                                                                                       
\multicolumn{16}{l}{\footnotesize {\bf Table 5b.3.} Undetected objects 
                    -- basic data-- {\it continued}}\\                                                                                         
\smallskip \\                                                                                                                                                                                           
\hline                                                                                                                                                                                                  
\vspace{-2 mm} \\                                                                                                                                                                                       
No & $K_{20}$ & $J-K$ & $H-K$ & $\mu_{K5}$ & $b/a$ & $r_{K_{20}}$ & $B_{T_{c}}$ & $D_{25}$ & $\mu_{B_{25}}$ & $T$ & rms &                                                                                          
 $I_{HI}$ &    \MHILKfr\  &   \MHILBfr\  & search \\                                                                                                                                                         
    & mag & mag & mag & $\frac{mag}{''^2}$  &  & $''$ & mag & $'$ & $\frac{mag}{''^2}$ &   & mJy & Jy km/s   &   &  & mode \\
\vspace{-2 mm} \\                                                                                                                                                                                       
\hline                                                                                                                                                                                                  
\vspace{-2 mm} \\
\end{tabular}                                                                                                                                                                                           
}                                                                                                                                                                                                       
\normalsize                                                                                                                                                                                             
\end{sidewaystable}

\begin{sidewaystable}                                                                                                                                                                                   
\centering                                                                                                                                                                                              
\bigskip                                                                                                                                                                                                
{\scriptsize                                                                                                                                                                                            
\begin{tabular}{lrrrrrrrrrrccccl}                                                                                                                                                                       
\multicolumn{16}{l}{\footnotesize {\bf Table 5b.4.} Undetected objects 
                                   -- basic data - {\it continued}}\\                                                                                         
\smallskip \\                                                                                                                                                                                           
\hline                                                                                                                                                                                                  
\vspace{-2 mm} \\                                                                                                                                                                                       
No & $K_{20}$ & $J-K$ & $H-K$ & $\mu_{K5}$ & $b/a$ & $r_{K_{20}}$ & $B_{T_{c}}$ & $D_{25}$ & $mu_{B_{25}}$ & $T$ & rms &          
 $I_{HI}$ &    \MHILKfr\  &   \MHILBfr\  & search \\   
    & mag & mag & mag & $\frac{mag}{''^2}$  &  & $''$ & mag & $'$ & $\frac{mag}{''^2}$ &   & mJy &          
 Jy.km/s   &   &  & mode \\                                                                                                                                                                                   
\vspace{-2 mm} \\                                                                                                                                                                                       
\hline                                                                                                                                                                                                  
\vspace{-2 mm} \\
\end{tabular}                                                                                                                                                                                           
}                                                                                                                                                                                                       
\normalsize                                                                                                                                                                                             
\end{sidewaystable}

%
\begin{sidewaystable}                                                                                                                                                                                   
\centering                                                                                                                                                                                              
\bigskip                                                                                                                                                                                                
{\scriptsize                                                                                                                                                                                            
\begin{tabular}{lrrrrrrrrrrccccl}                                                                                                                                                                       
\multicolumn{16}{l}{\footnotesize {\bf Table 5b.5.} Undetected objects 
                                   -- basic data - {\it continued}}\\                                                                                         
\smallskip \\                                                                                                                                                                                           
\hline                                                                                                                                                                                                  
\vspace{-2 mm} \\                                                                                                                                                                                       
No & $K_{20}$ & $J-K$ & $H-K$ & $\mu_{K5}$ & $b/a$ & $r_{K_{20}}$ & $B_{T_{c}}$ & $D_{25}$ & $mu_{B_{25}}$ & $T$ & rms &          
 $I_{HI}$ &    \MHILKfr\  &   \MHILBfr\  & search \\   
    & mag & mag & mag & $\frac{mag}{''^2}$  &  & $''$ & mag & $'$ & $\frac{mag}{''^2}$ &   & mJy &          
 Jy.km/s   &   &  & mode \\
\vspace{-2 mm} \\                                                                                                                                                                                       
\hline                                                                                                                                                                                                  
\vspace{-2 mm} \\
\end{tabular}                                                                                                                                                                                           
}                                                                                                                                                                                                       
\normalsize                                                                                                                                                                                             
\end{sidewaystable}

\newpage
\clearpage

\begin{figure*} 
\centering
\includegraphics[width=15cm]{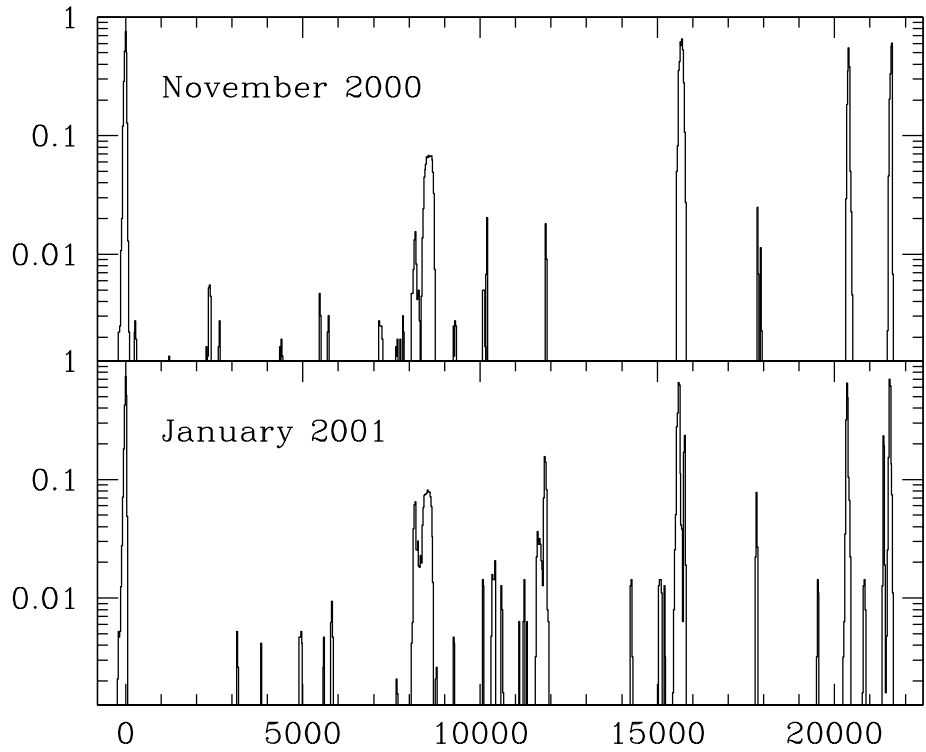}
\caption [] {Plots showing the relative occurrence of radio 
frequency interference (RFI) during the two observing runs, for the observations
made in the velocity search mode (see Sect. 3.1).
It also shows the omnipresence of Galactic \HI\ emission around 0 \kms. 
Shown as function of radial velocity is the relative number 
of spectra with signals having a flux density exceeding the 20 mJy level, in
25 \kms\ wide bins.
}
\end{figure*}

%
\begin{figure*} 
\centering
\includegraphics{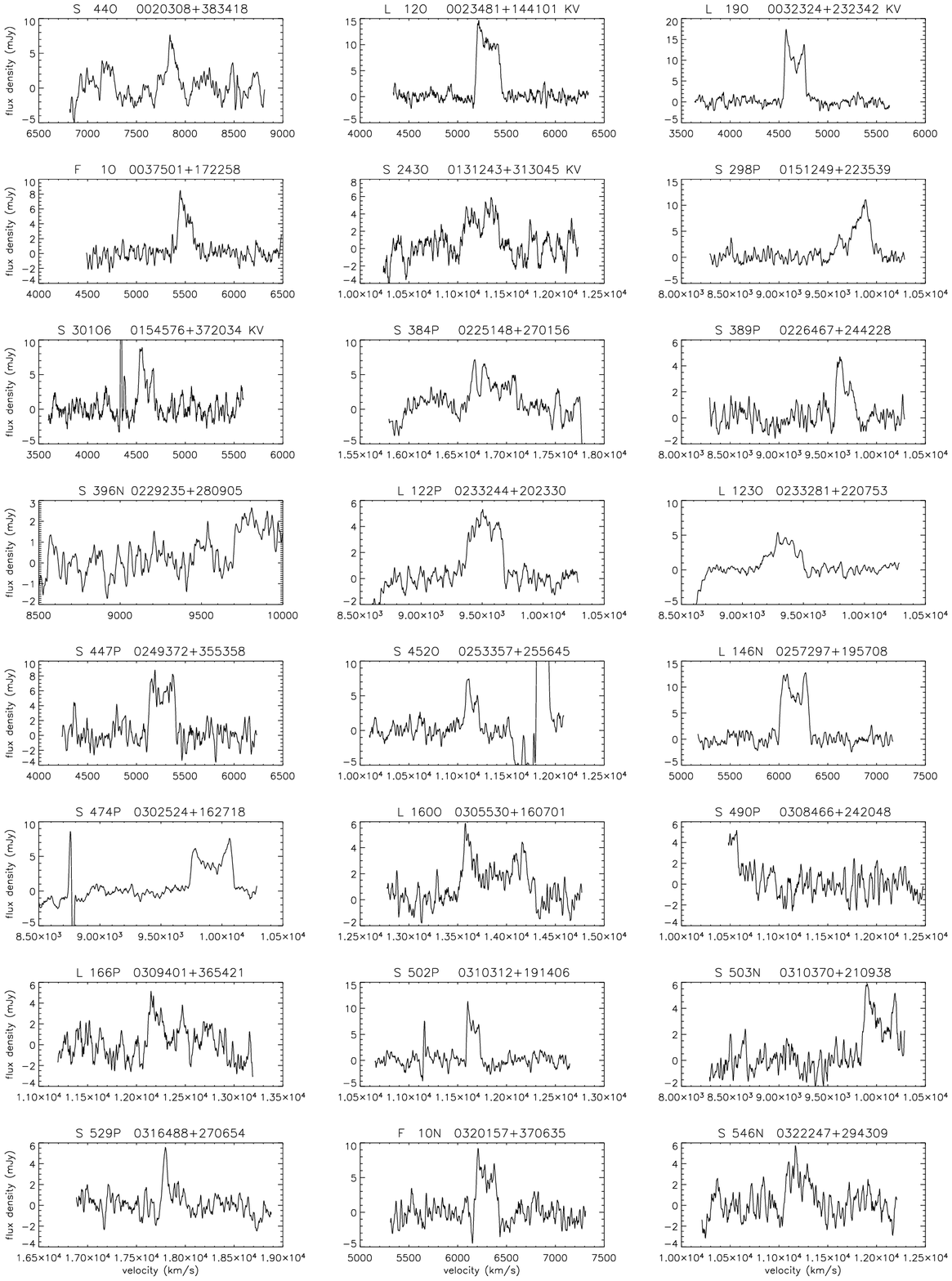}
\caption [] {{\bf a.}\, Arecibo 21-cm \HI\ line spectra of the clearly detected objects (see Table 3). 
Velocity resolution is 14.3 \kms\ (velocity search mode) and 16.4 \kms\ (known velocity mode), 
radial heliocentric velocities are according to the optical convention. Galaxies with previously know
redshifts, detected in the `known velocity' mode, are indicated by the designation `KV' following 
their coordinates in the header of their spectrum.
}
\end{figure*}

\begin{figure*} 
\centering
\addtocounter{figure}{-1}
\includegraphics{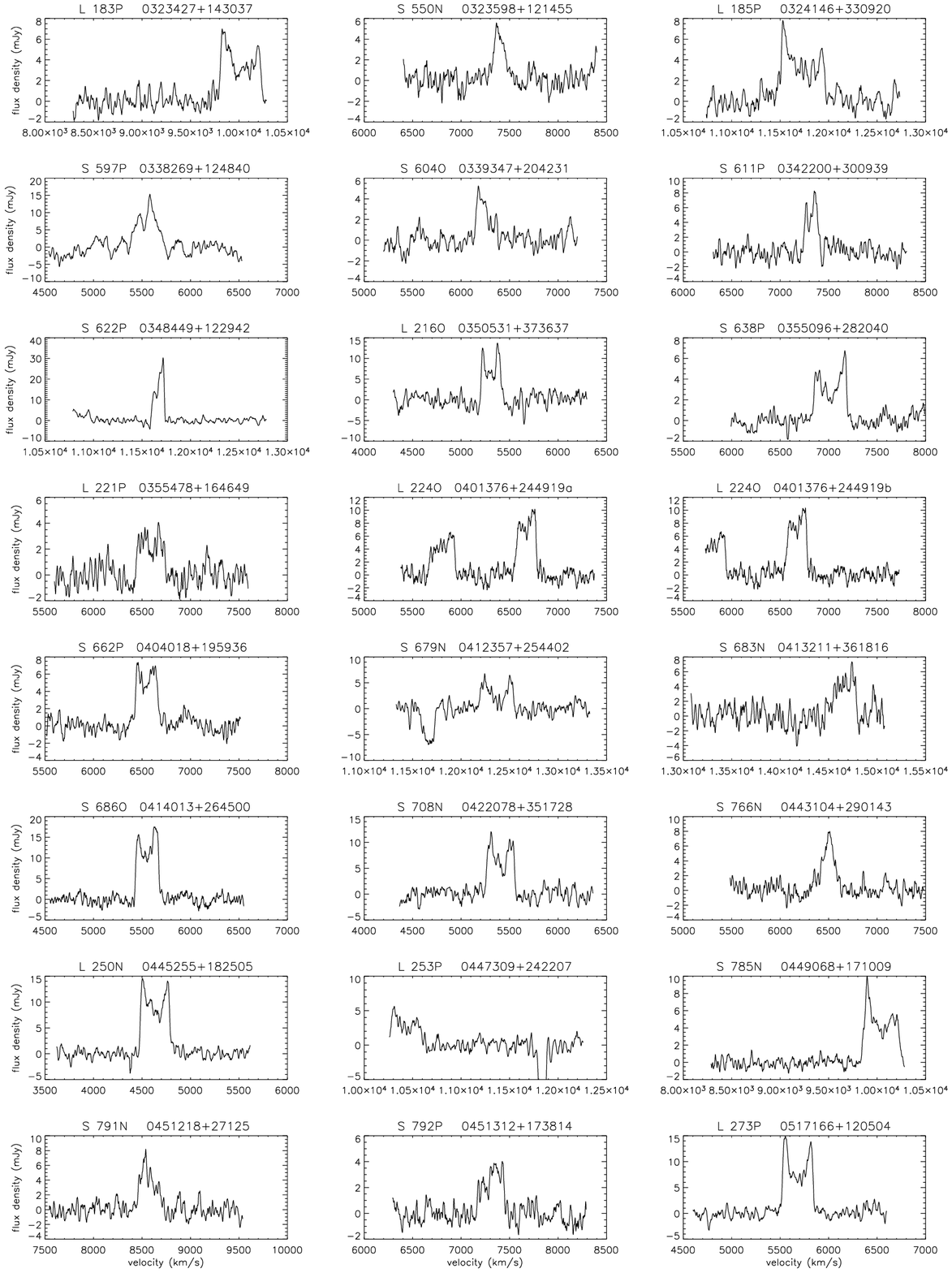}
\caption [] {{\bf b.}\, Arecibo 21-cm \HI\ line spectra of clearly detected objects {\it - continued}. 
}
\end{figure*}

\begin{figure*} 
\centering
\addtocounter{figure}{-1}
\includegraphics{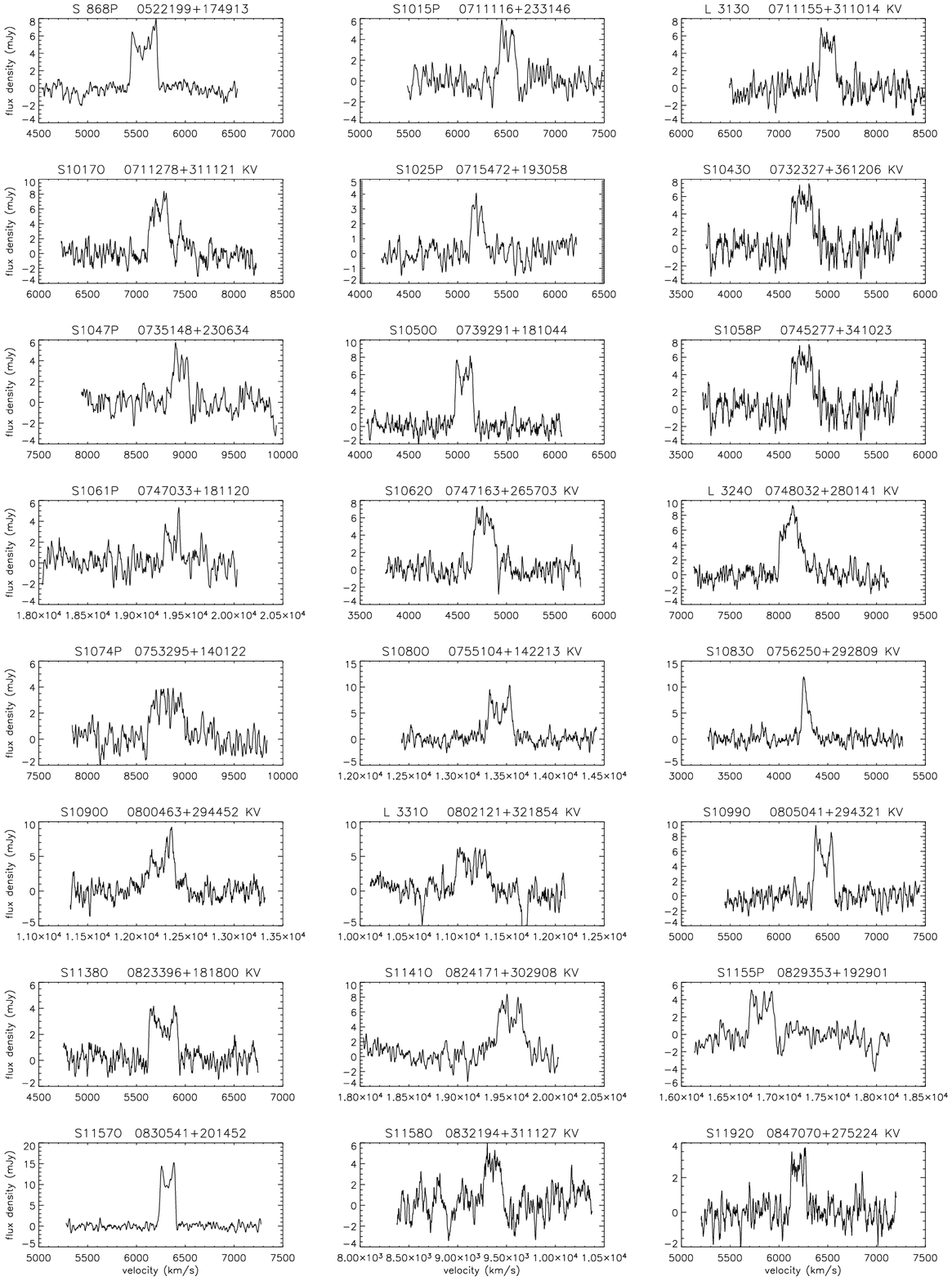}
\caption [] {{\bf c.}\, Arecibo 21-cm \HI\ line spectra of clearly detected objects {\it - continued}. 

}
\end{figure*}

\begin{figure*} 
\centering
\addtocounter{figure}{-1}
\includegraphics{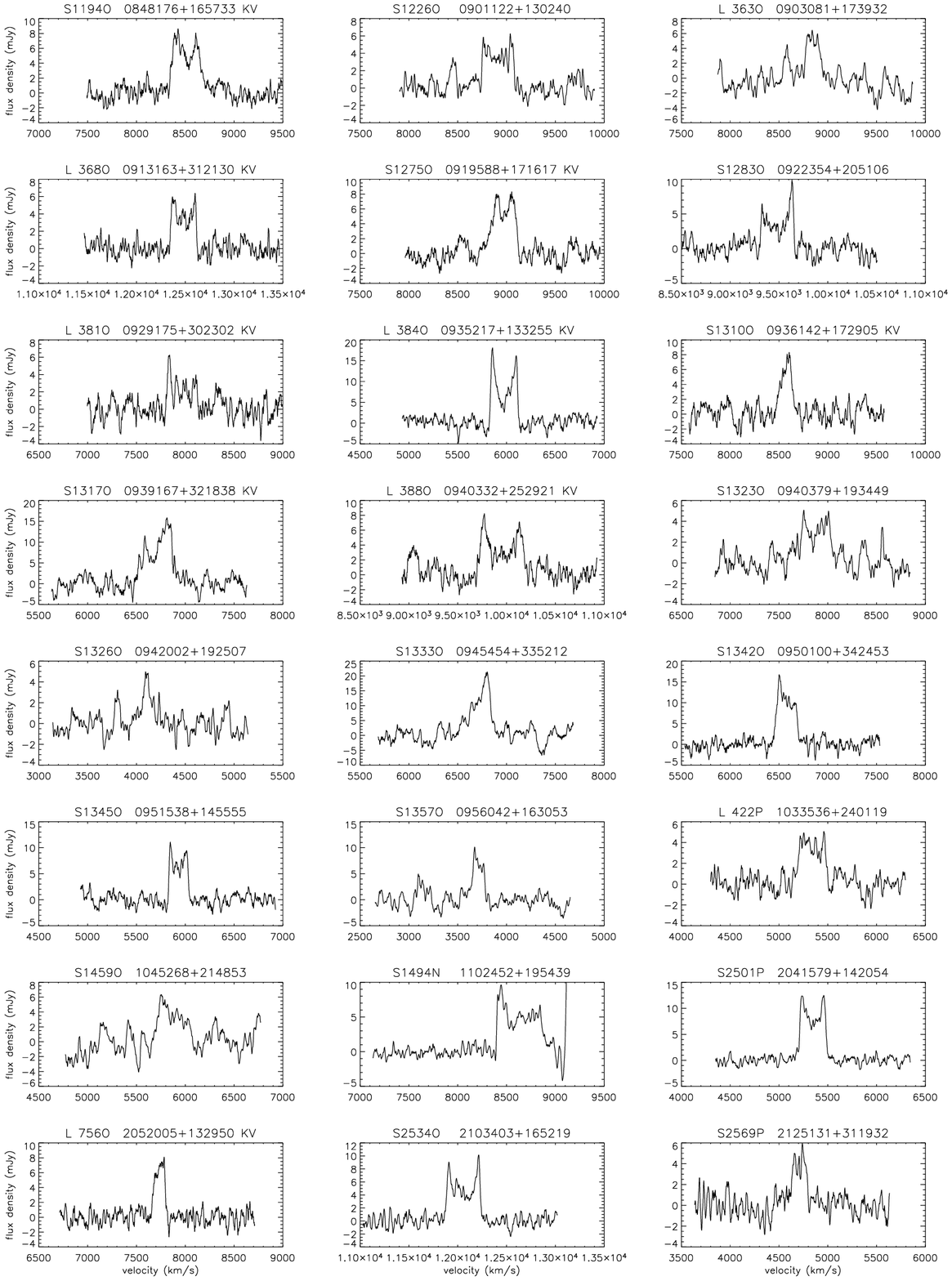}
\caption [] {{\bf d.}\, Arecibo 21-cm \HI\ line spectra of clearly detected objects {\it - continued}. 
}
\end{figure*}

\begin{figure*} --- 2e -----------------
\centering
\addtocounter{figure}{-1}
\includegraphics{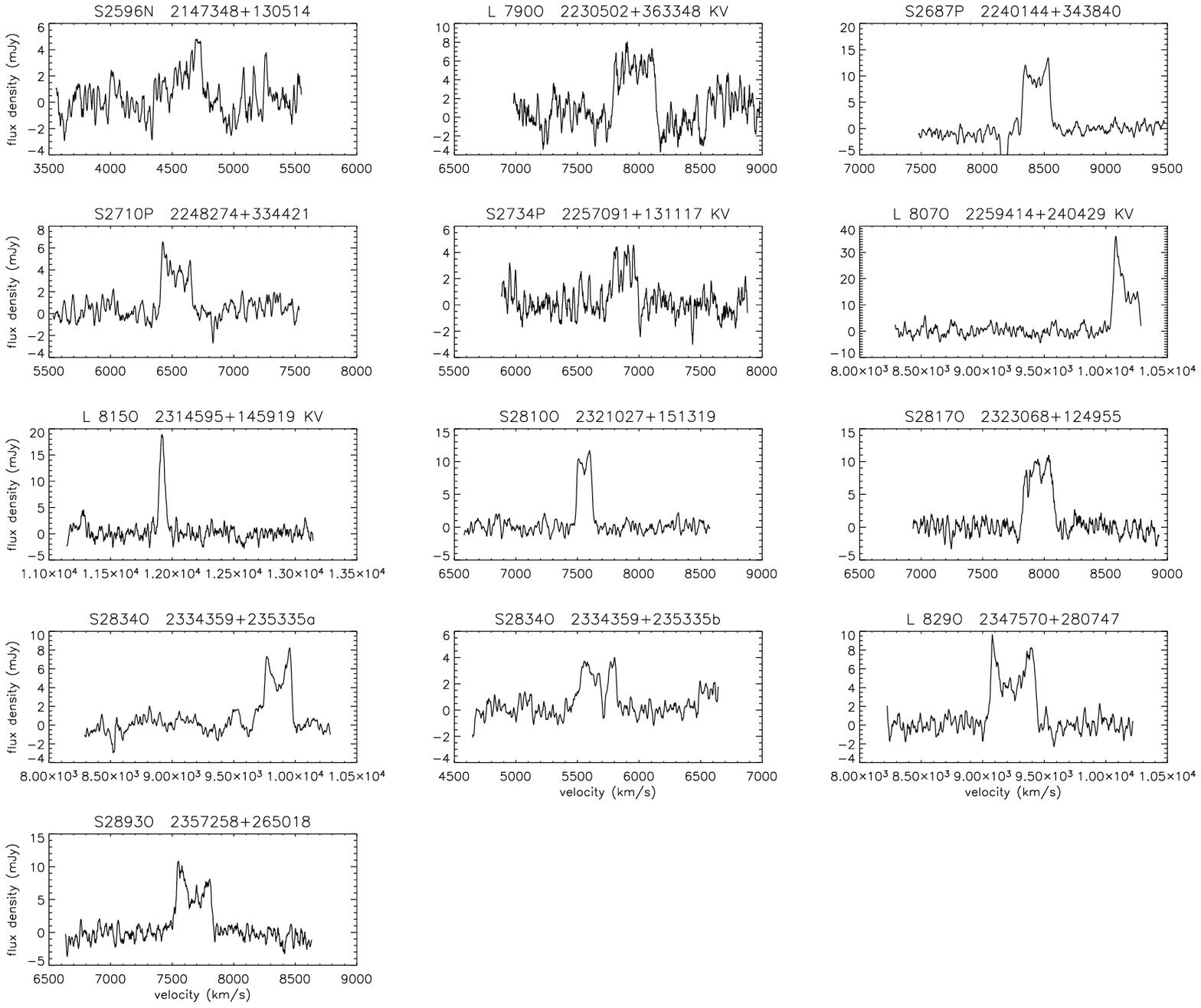}
\vspace{-8cm}
\caption [] {{\bf e.}\, Arecibo 21-cm \HI\ line spectra of clearly detected objects {\it - continued}. 
}
\end{figure*}

\begin{figure*} --- 3 -----------------
\centering
\includegraphics{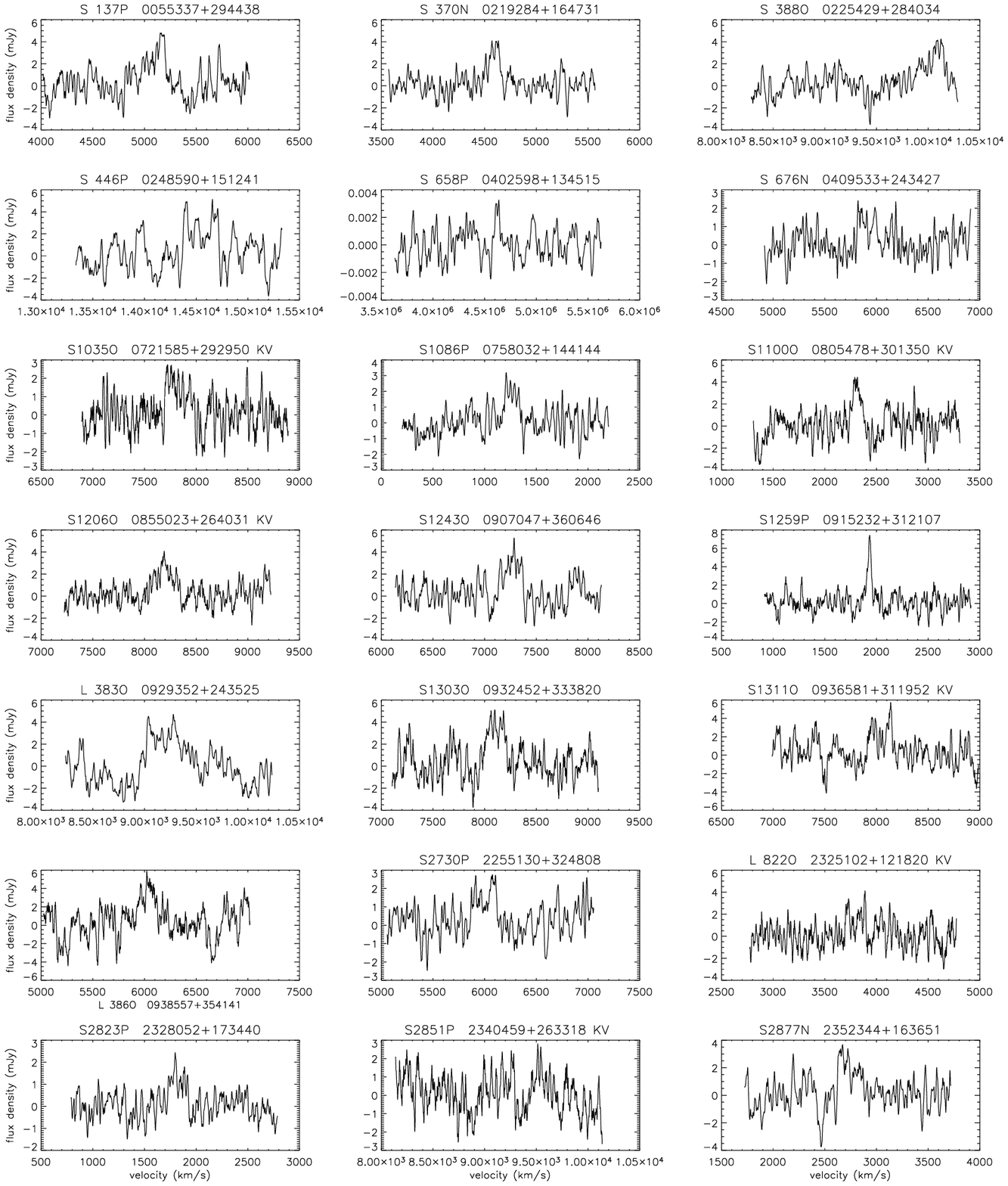}
\vspace{-3cm}
\caption [] {Arecibo 21-cm \HI\ line spectra
of marginal detections (see Table 4). 
Velocity resolution is 14.3 \kms\ (velocity search mode) and 16.4 \kms\ 
(known velocity mode), radial heliocentric velocities are according to the optical 
convention. Galaxies with previously know redshifts, detected in the 'known velocity' mode,
are indicated by the designation 'KV' following their coordinates in the header of their spectrum.
}
\end{figure*}

\begin{figure*}
\includegraphics[width=10cm,angle=90]{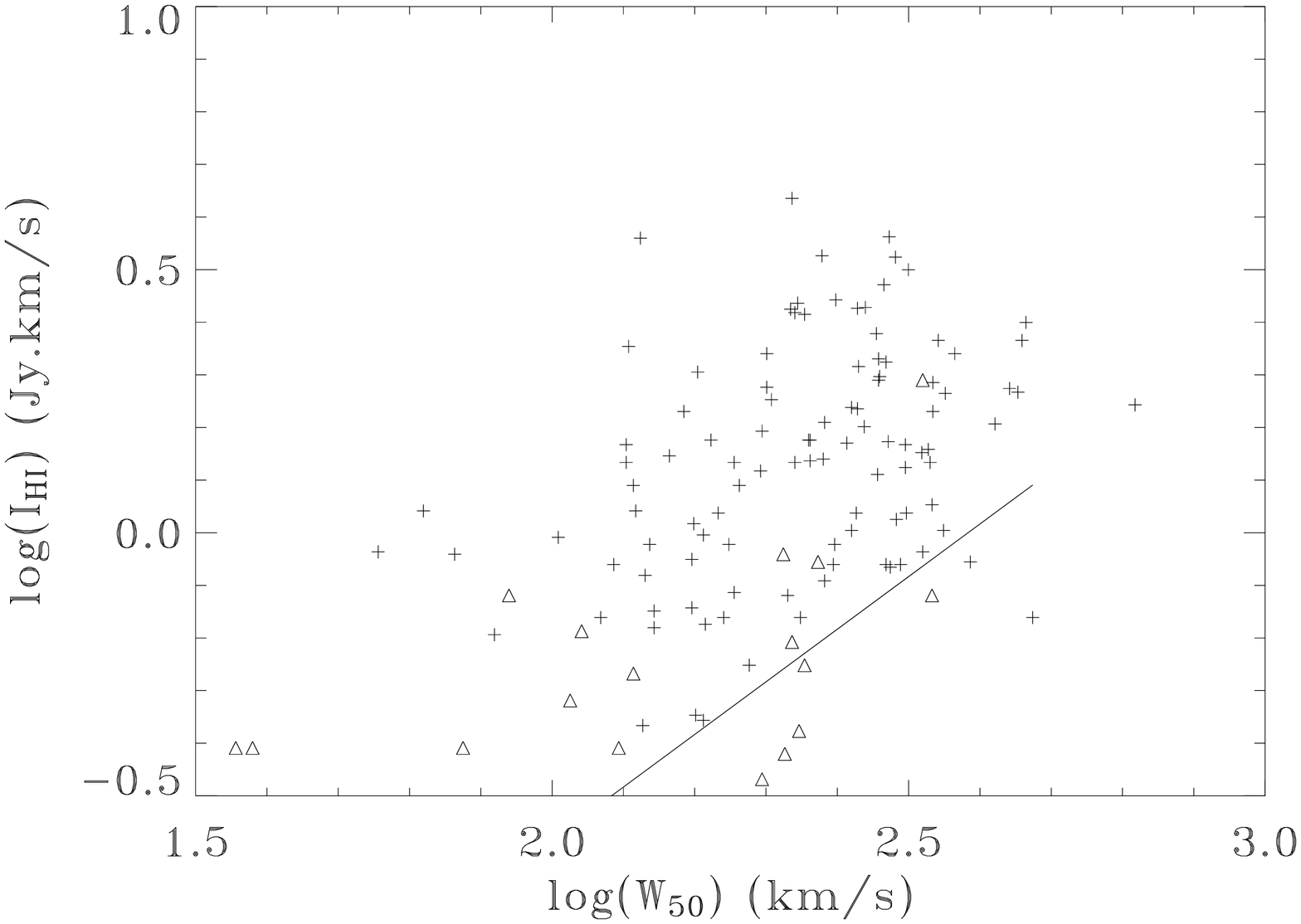}
\caption [] {Distribution of integrated \HI\ line fluxes ($I_{HI}$) as a function of the \HI\ line
FWHM, $W_{50}$.
The straight line indicates the 3$\sigma$ detection limit for a 250 \kms\ wide, 
flat-topped spectral line, based on the average rms noise level of the data. 
Clear detections are represented by crosses, marginal detections by triangles.
}
\end{figure*}
\end{document}